\documentclass[apj,tighten,iop,twocolumn]{emulateapj}
\usepackage{bm}
\newcommand{\lunits}{erg~s$^{-1}$}

\newcommand{\mgal}{$M_{\rm gal}$}
\newcommand{\mstar}{$M_{\ast}$}
\newcommand{\msun}{$M_{\odot}$}
\newcommand{\msuny}{\msun yr$^{-1}$}

\newcommand{\aco}{$\alpha_{\rm CE}$}
\newcommand{\acol}{$\lambda\alpha_{\rm CE}$}

\newcommand{\ewind}{$\eta_{\rm wind}$}

\newcommand{\kdcbh}{$\kappa_{\rm DCBH}$}

\newcommand{\lpair}{$\cal L_{\rm pair}$}

\newcommand{\lpairkm}{${\cal L}_{{\rm pair,} km}$}
\newcommand{\lapairn}{$\Lambda_{\rm n, pair}$}
\newcommand{\laglobn}{$\Lambda_{\rm n, global}$}
\newcommand{\lglob}{$\cal L_{\rm global}$}

\newcommand{\lx}{$L_X$}
\newcommand{\lxtt}{$L_{X,{\rm 0.3-10.0 keV}}$}

\newcommand{\lhx}{$L_{\rm HX}$}

\newcommand{\pold}{\lq\lq old\rq\rq}     
\newcommand{\pyoung}{\lq\lq young\rq\rq} 

\newcommand{\x}{X-ray}

\newcommand{\chandra}{{\it Chandra}}

\newcommand{\spitzer}{{\it Spitzer}}
\newcommand{\stk}{{\tt StarTrack}}

\newcommand{\wav}{{\sc wavdetect}}
\newcommand{\ace}{{\sc ae}}

\newcommand{\er}{Equation~\ref}
\newcommand{\fr}{Fig.~\ref}
\newcommand{\scr}{Sec.~\ref}
\newcommand{\tr}{Table~\ref}

\newcommand{\exi}{\begin{equation}}
\newcommand{\exo}{\end{equation}}

\newcommand{\ten}[2]{$#1\times 10^{#2}$} 

\slugcomment{Accepted by ApJ, June 20, 2013}

\shorttitle{Modeling X-ray binaries in galaxies}
\shortauthors{Tzanavaris et al.}

\begin{document}

\title{Modeling X-ray binary evolution in normal galaxies: Insights from SINGS}


\author{
P.~Tzanavaris\altaffilmark{1,2,3},
T.~Fragos\altaffilmark{4,5}, 
M.~Tremmel\altaffilmark{6},
L.~Jenkins\altaffilmark{1},
A.~Zezas\altaffilmark{7,8,4},
B.~D.~Lehmer\altaffilmark{1,2,9},
A.~Hornschemeier\altaffilmark{1},
V.~Kalogera\altaffilmark{10},
A.~Ptak\altaffilmark{1},
A.~R.~Basu-Zych\altaffilmark{1}
}


\altaffiltext{1}{Laboratory for X-ray Astrophysics, 
NASA/Goddard Spaceflight Center,Mail Code 662, Greenbelt, Maryland, 20771, USA}
\altaffiltext{2}{Department of Physics and Astronomy,
The Johns Hopkins University, Baltimore, MD 21218, USA}
\altaffiltext{3}{NPP Fellow}
\altaffiltext{4}{Harvard-Smithsonian Center for Astrophysics,
60 Garden Street, Cambridge, MA 02139, USA}
\altaffiltext{5}{CfA and ITC prize Fellow} 
\altaffiltext{6}{Department of Astronomy,  University of Washington,  Box
351580, U.W., Seattle, WA 98195-1580, USA}
\altaffiltext{7}{Department of Physics, University of Crete, P.O. Box 2208,
71003 Heraklion, Crete, Greece}
\altaffiltext{8}{IESL,  Foundation  for  Research  and  Technology,   71110
Heraklion, Crete, Greece}
\altaffiltext{9}{Einstein Fellow}
\altaffiltext{10}{Center   for   Interdisclipinary   Research   and   Exploration in  Astrophysics  and  Department  of  Physics  and  Astronomy,
Northwestern  University,  2145  Sheridan  Road,  Evanston,  IL
60208, USA}


\begin{abstract}

We present the largest-scale comparison to date between
observed extragalactic \x\ binary (XRB) populations and theoretical
models of their production.  We construct observational X-ray
luminosity functions (oXLFs) using \chandra\ observations of 12
late-type galaxies from the Spitzer Infrared Nearby Galaxy
Survey (SINGS). 
For each galaxy, we obtain theoretical XLFs (tXLFs)
by combining XRB synthetic models, constructed with
the population synthesis code \stk,
with observational star formation histories
(SFHs).
We identify highest-likelihood
models both for individual galaxies and globally, averaged over the full galaxy
sample. Individual tXLFs successfully reproduce
about half of oXLFs,
but for some galaxies we are unable to find
underlying source populations, indicating that galaxy
SFHs and metallicities are not well matched and/or
XRB modeling requires calibration on larger observational samples.
Given these limitations, we find
that best models are consistent with 
a product of common envelope ejection efficiency and central
donor concentration $\simeq 0.1$,
and a 50\%\ uniform -- 50\%\ \lq\lq twins\rq\rq\ initial mass-ratio distribution.
We present and discuss constituent subpopulations of tXLFs
according to donor, accretor and stellar population characteristics.
The galaxy-wide \x\ luminosity due to low-mass and high-mass XRBs, 
estimated via our
best global model tXLF,
follows the general trend
expected from the \lx\ - star formation rate and \lx\ - stellar mass
relations of Lehmer et al. (2010).  Our best models are also in
agreement with modeling of the evolution both of
XRBs over cosmic time and of the 
galaxy \x\ luminosity with redshift.
\end{abstract}

\keywords{binaries: close — galaxies: spiral — stars: evolution — \x s: binaries}

\section{Introduction} 
Binary stars constitute a substantial fraction of stellar populations
(SPs). In galactic fields and low density agglomerations such as open
clusters between $\sim 40$\%\ and $\sim 75$\%\ of stars are binaries
\citep{duquennoy1991,fischer1992,fan1996,raghavan2010,sana2012}. In
fact, stellar binarity may well be a universal characteristic of
stellar evolution, since many single stars may either have been
through a binary phase, only to be ejected later, or be the result of
a binary merger \citep{sana2012}. Compared to single stars, binary
stars are hosts to a range of additional processes (mass and angular
momentum transfer, wind accretion, Roche-lobe overflow, common
envelope ejection etc.), making them ideal astrophysical laboratories
for a whole range of physics not represented among single stars. Some
of the most interesting processes that can be probed are associated
with accretion onto primaries that are compact objects (neutron stars
and black holes). Due to the extreme energies involved, such cases are
observationally identified as X-ray binaries (XRBs).

There have been many observational studies of XRB populations in {\it
  external} galaxies. Earlier work with the {\it Einstein} satellite
showed that the X-ray emission is dominated by XRBs with low-mass
(LMXBs) or high-mass donors (HMXBs) in early and late-type galaxies,
respectively \citep{kim1992}. Using {\it Chandra}'s sub-arcsecond
resolution, this result has now been established for individually
detected XRBs in nearby galaxies \citep[e.g.][see also
  \citet{fabbiano2006} and references
  therein]{kong2002,soria2002,trudolyubov2002,sivakoff2003,kim2003,gilfanov2004a,gilfanov2004b,kim2004,zhang2012,binder2012,binder2013}. Due
to the longer evolutionary timescales of LMXBs, their integrated X-ray
emission is an indicator of a galaxy's total stellar mass
\citep{gilfanov2004b,bogdan2010,zhang2011,boroson2011}. In contrast,
relatively short-lived HMXBs probe a galaxy's star formation rate
\citep[(SFR),][]{grimm2003,ranalli2003,gilfanov2004a,persic2007,shtykovskiy2007,lehmer2010,mineo2012}. These
observations have also established a break in the XLF of LMXBs at low
luminosities \citep[$L_X\sim 10^{37}$~\lunits,][]{gilfanov2004b,revnivtsev2008,revnivtsev2011} both in the
Milky Way and in external galaxies. 

In the past, semi-analytical theoretical models have been introduced
for the study of XRB populations. \citet{white1998} and \citet{ghosh2001}, assumed a
time-dependent SFR and a simple rate model to study the evolution of
an arbitrary XRB population. These models predicted that the time
required for binaries to reach the X-ray phase leads to a significant
time delay between a star-formation episode and the production of
X-ray emission from X-ray binaries from the
population. \citet{wu2001} created a simple birth-death model, in
which the lifetimes of the binaries are inversely proportional to
their X-ray luminosity, and calculated the XLFs of spiral
galaxies. His models reproduce some features, such as the luminosity
break in the observed XLFs of spiral galaxies. \citet{piro2002} argued
that the majority of LMXBs in the field of elliptical galaxies have
red giant donors feeding a thermally unstable disk and stay in this
transient phase for at least 75\% of their life. Most recently, in a
series of papers, \citet{bhadkamkar2012,bhadkamkar2013a,bhadkamkar2013b} 
started from standard distributions of the
parameters of those primordial binaries which are the progenitors of
XRBs, and followed the transformation of these distributions with the
aid of a Jacobian formalism as the binaries progress through different
evolutionary phases. Following this methodology, they were able to
derive estimates for the XLF and other population statistical
properties, for both HMXBs and LMXBs.

Binary population synthesis (PS) modeling codes can provide a unique
tool to understand the physical properties of this important
population in a statistical sense. \citet{lipunov1996} used for the
first time a PS code to investigate the evolution of XRBs in the
central part of the Milky Way over the course of 10 My \citep[see][for
  a description of their code ``Scenario Machine'']{lipunov2009}. A
number of other codes have been developed mostly within the last
decade or so \citep{hurley2002,kiel2006,belczynski2008}. These codes
have been used by several authors to carry out a variety of
investigations, including the study of double compact object mergers
\citep[e.g.][]{belczynski2002b}, the formation of ultrashort XRBs
\citep[e.g.][]{belczynski2004a}, the numbers and spatial distributions
of XRBs in star clusters \citep[e.g.][]{sepinsky2005}, the evolution
of XRBs in a brief star-formation episode as a model of starburst
systems \citep{eracleous2006}, the formation of binary millisecond
pulsars \citep[e.g.][]{pfahl2003,hurley2010}, binary fractions in
globular clusters \citep[e.g.][]{ivanova2005,hurley2007}, and numbers
and birthrates of symbiotic XRBs in the Galaxy \citep[e.g.][]{lu2012}.

Physically motivated PS modeling and detailed comparisons of XLF
characteristics can be used to understand how these observations are
linked to the formation and evolution of XRB populations in
galaxies. This type of work was pioneered by \citet{belczynski2004a}
who compared a theoretical X-ray luminosity function (tXLF) with
observational X-ray luminosity functions (oXLFs) for XRBs in the dwarf
irregular galaxy NGC 1569 obtained with {\it
  Chandra}. \citet{linden2009,linden2010} studied the XLF for HMXBs
and Be XRBs in the SMC. \citet{fragos2008,fragos2009} modeled the XLFs
in the two elliptical galaxies NGC 3379 and NGC 4278, and investigated
the contributions from subpopulations of LMXBs. \citet{zuo2011} used
the PS code of \citet{hurley2002} to investigate the X-ray-evolution
of late-type galaxies over $\sim 14$ Gy of cosmic time.

To carry out this type of work, it is necessary to combine PS models
with star formation history (SFH) information for specific
galaxies. For their elliptical galaxies \citet[][hereafter
  F08]{fragos2008} assumed an initial $\delta$-function star formation
episode. For later type galaxies SFH can be obtained via spectral
energy distribution (SED) modeling, which requires multiwavelength
information for a given galaxy.

In this paper we use a sample of 12 nearby galaxies from the Spitzer
Infrared Nearby Galaxy Survey (SINGS) covering a range in star forming
properties to extend binary PS modeling to later type systems. We
construct oXLFs for their off-nuclear point source XRB populations and
tXLFs by combining population synthesis modeling results, obtained by
means of the PS code {\tt StarTrack}, and star formation histories from
the literature. This allows us to compare the two sets of XRB XLFs and
investigate the range of acceptable values for XRB formation and
evolution parameters.

This paper is part of a larger effort to understand the formation and
evolution of extragalactic XRBs by means of the most advanced PS
modeling to date. Other papers in the series include \citet[][hereafter F13]{fragos2013} and 
Tremmel et al. (2013,
accepted, astroph/1210.7185, hereafter T13). F13 study the
evolution of the global XRB population with redshift by using the
Millennium-II simulation as initial conditions. They accurately
reproduce local group HMXB and LMXB luminosity scaling relations with
SFR and $M_{*}$ \citep{lehmer2010,mineo2012}, respectively. T13 use
the same initial conditions to explain observational XLFs for the
integrated XRB emission from entire galaxies \citep{tzanavaris2008}
and make predictions for higher redshifts. In this paper we apply the
same grid of PS models, combining it with SFH information for nearby
galaxies \citep{noll2009}.

The structure of the paper is as follows: In Section 2 we present the
observational sample and the construction of oXLFs. Section 3
discusses the calculation of tXLFs. Section 4 presents likelihood
functions for establishing best PS models. Results are presented and
discussed in Section 5. Section 6 gives a summary and discusses future
prospects.

\section{Observational Sample and XLFs}\label{sec_obs}
We use galaxies selected from the Spitzer Infrared
Nearby Galaxy Survey \citep[SINGS,][]{kennicutt2003}. This survey was
designed to be a diverse sample of intrinsic galaxy properties with
multiwavelength data ranging from the ultraviolet to the far infrared.
As part of a large \chandra\ program \citep[XSINGS,][]{jenkins2010},
the Advanced CCD Imaging Spectrometer (ACIS) was used to extend the
survey's wavelength coverage to the \x\ regime.  Details regarding the
sample selection, \x\ observations, source detection and characterization
will be presented in a forthcoming publication (Jenkins et al.,
in prep.)  Briefly, basic \x\ data reduction was carried out using
standard \chandra\ \x\ Center tools. Point source detection was performed in
the soft (0.3-2.0 keV), hard (2.0-10.0 keV) and full (0.3-10.0 kev)
band with CIAO\footnote{http://cxc.harvard.edu/ciao} \wav\ to
construct a candidate source list. The final list was produced by
using the software {\sc acis extract} \citep[\ace,][]{broos2010} to
perform aperture photometry and produce a catalog of point sources
with associated fluxes and luminosities for each galaxy.

We use a sub-sample of 12 SINGS galaxies that have 
SFHs from spectral energy distribution
fitting with the code
CIGALE\footnote{http://cigale.oamp.fr/} \citep{noll2009}.  Galaxies
are also selected to have at least 15 detected, non-nuclear, \x\ point
sources as a prerequisite for the production of meaningful XRB
XLFs. Details of the galaxy sample are given in
\tr{tab-sample}. \spitzer-infrared and \chandra-\x\ galaxy images are
shown in \fr{fig_sample}.  To illustrate the properties of our sample
relative to the rest of the SINGS galaxies, in \fr{fig_cmsm} we show both a
color-magnitude diagram and a plot of star formation rate vs. stellar
mass for the full SINGS sample, highlighting our galaxies. In both
plots, galaxies are separated into three broad morphological
categories, namely E/S0 (shown in red), Sa-Sbc (magenta), and
Sc-Im (blue). Compared to the rest of the SINGS galaxies, our
systems have intermediate to high SFR and \mstar, and intermediate
to red colors.

At low point source luminosities incompleteness effects arise, 
compromising the construction of unbiased XLFs. These can
be mitigated either by limiting observational XLFs to the luminosity
range in which incompleteness is not significant or by performing
incompleteness corrections.  The second approach is preferable since
it allows the construction of XLFs covering a wider dynamic range in
\x\ luminosity. We use the method of \citet[][see also
  \citet{zezas2002}]{zezas2007} to create simulated \x\ source
catalogs, calculate the source detection probabilities and obtain
incompleteness corrections as a function of source and background
intensity (in counts) and off-axis angle for \x\ detected sources.
Note that this detection probability is otherwise independent of other
source characteristics such as detection band and intrinsic source
spectrum.

oXLFs are shown in \fr{fig_otxlf} by the blue curves
before (dotted) and after (solid) completeness correction.
In practice, we find that completeness corrections
are small and do not change the models that are associated
with the highest-likelihood tXLF for
a given galaxy. Specifically, a quality check on our
best populated galaxy NGC 1291, for which we 
apply both methods, shows that our conclusions do not change.
We note that the highest-likelihood tXLF is also shown
in \fr{fig_otxlf}
(in red and dark grey; see \scr{sec_lik} for details).

\section{Population Synthesis Modeling}
\subsection{\stk}\label{sec_stk}

The main tool we use to perform our PS simulations is
\stk, a state-of-the-art PS code that
has been tested and calibrated using detailed mass transfer star
calculations and observations of binary populations. \stk\
has been applied to numerous interpretation studies of X-ray and radio
pulsar binary populations, as well as $\gamma$-ray bursts and binary
black holes in the context of gravitational-wave sources \citep[see detailed description in][and references therein]{belczynski2008}. In
summary the code incorporates all the important physical processes of
binary evolution:
 
\noindent\textit{(i)} The evolution of single stars and
non-interacting binary components from zero-age main sequence
to remnant formation is
followed with the use of high-quality analytic formulae
\citep{hurley2000}. Various wind mass loss rates dependent on stellar
evolutionary stage are incorporated and their effect on stellar
evolution is taken into account.

\noindent\textit{(ii)} Changes in all the orbital properties are
tracked. Through numerical integration of four differential equations,
the evolution of orbital separation, eccentricity and component spins
is tracked; these depend on tidal interactions as well as angular
momentum losses associated with magnetic braking, gravitational
radiation and stellar wind mass losses.

\noindent\textit{(iii)} All types of mass-transfer phases are
calculated: Stable, driven by nuclear evolution or angular momentum
loss, and thermally or dynamically unstable.

\noindent\textit{(iv)} Supernova explosions are treated accounting for
mass loss and asymmetries with natal kicks to neutron stars and black
holes at birth; systemic velocities for all binaries are calculated.

\noindent\textit{(v)} The calculation of mass transfer rates in
binaries with accreting neutron stars and black holes (driven by
stellar winds or Roche-lobe overflow) has been calibrated against
detailed mass transfer sequences, and X-ray luminosities are calculated
by incorporating appropriate band-pass corrections and spectral models.

This paper uses results that are based on a recent major revision of the \stk\
code that includes updated stellar wind prescriptions
and their re-calibrated dependence on metallicity
\citep{belczynski2010}. Two newer updates have not been
taken into account, as our simulations were complete before these
updates had been implemented. For reference, these are (1) a
revised neutron star and black hole mass spectrum, leading to
fully consistent supernova simulations
\citep{belczynski2012,fryer2012}; and
(2) a more physical treatment of donor stars in common envelopes 
via actual $\lambda$ values \citep{dominik2012}, where
$\lambda$ is a measure of the
donor's central concentration and the envelope binding energy.

\tr{tab-param} lists the full set of input parameters used in our PS modeling.
We construct a large PS model grid by using a range of values for
these parameters as indicated in the table.
For detailed discussion of how each parameter
affects the overall XRB population, we refer the reader to
F13 (Section 5.1 and Fig.~6) and T13 (Section 4.3 and Fig.~7).

The input parameters fall into two categories. First,
parameters that mostly characterize initial properties of the population,
such as initial mass function (IMF), initial binary mass ratio ($q$),
and distribution of initial orbital separations.
For these parameters we usually have information from observational
surveys of binary stars, constraining the range of values used.
The second category comprises parameters that are associated
with physical processes that are poorly understood, such as the
efficiency, \aco, of converting orbital into thermal energy that will be
used to expel the donor's envelope during a common envelope (CE) phase.
In this work the related parameter varied is \acol, which
is the product of this efficiency and the donor central concentration, $\lambda$.

The complete grid of 288 PS models is the same as that used by
F13 and T13. To construct this grid, we vary
all
parameters known from earlier studies
to affect XRB evolution and formation of compact objects
\citep{belczynski2007,fragos2008,fragos2009,linden2009,belczynski2010,fragos2010}. 
Specifically we use:
\begin{itemize}
\item Four \acol\ values (0.1, 0.2, 0.3, 0.5);
\item Three stellar wind strengths, \ewind\ (0.25, 1.0, 2.0). 
This parameter is used
to multiply the stellar wind prescription of \citet{belczynski2010};
\item The distribution of natal kicks for BHs formed through 
direct collapse (no kicks or 10\%\ of the \citet{hobbs2005}
distribution for NSs);
\item A CE-HG flag for systems with a donor
in the Hertzsprung gap, either allowing all possible common envelope events
or always imposing a merger \citep{belczynski2007};
\item Three distributions of binary initial mass ratios, $q \equiv M_{\rm secondary} 
/ M_{\rm primary}$. This distribution specifies the mass of
the secondary, whereas the mass of the primary is governed
by the IMF. For our full model grid, we use a uniform (flat) distribution, $q=0\rightarrow 1$,
a twins distribution, $q=0.9\rightarrow 1$, and a mixed distribution,
50\%\ uniform and 50\%\ twins. However, in this paper
we do not use the 96 twin $q$
models $97-192$, as F13 have already clearly shown that these are very
inadequate in reproducing observed XRB populations, as
they prevent the production of LMXBs altogether. This limits
the models used in this paper to those in the ranges $1-96$ and $193-288$, 
i.e. 192 models in total.
\end{itemize}

In addition, for each of the 288 models in our full grid, we also
use nine metallicities,
keeping all other parameters same.  In this
paper we only consider solar metallicity models, as the best estimates
of \citet{moustakas2010} using two different methods appear to straddle
solar metallicity for all SINGS galaxies
in our study.
The best fit SEDs that we convolve with \stk\ models to construct
tXLFs for individual galaxies also assume solar metallicities
\citep{noll2009}.

Each model follows the evolution of \ten{5.12}{6} stars over 14 Gyr.
Since we are only using models $1-96$, corresponding to a uniform $q$
distribution, and models $193-288$, corresponding to a 50\% -- 50\%
mixed $q$ distribution, in relevant figures we indicate these model
ranges for clarity. For reference, the full list of 288 models and
associated parameters can be found in Table 4 of F13.

\subsection{Theoretical \x\ luminosities}\label{sec_X}
Since our goal is to construct XRB XLFs, we need to identify all
binaries in our simulations that become XRBs and register their
\x\ luminosities as a function of time.  XRBs are mass-transferring
binary stellar systems with a compact object accretor, either a black
hole (BH) or a neutron star (NS). 
Systems with donors less massive
than 3\msun\ are labeled LMXBs, and vice versa for HMXBs.
LMXBs are always Roche lobe overflowing (RLOF) systems, while HMXBs
are usually wind-fed, but can also exhibit RLOF behavior.  According to
whether they undergo thermal disk instability or not, RLOF systems can
be either transient or persistent, while wind-fed systems are always
persistent.
\footnote{Be XRBs are a special case, since, although they
are wind-fed, they show quasi-periodic outbursts. However,
the origin of this behavior is not the
thermal disk instability.}

We follow the methodology
described in F08 and \citet{fragos2009} to
identify all model XRB sources 
and keep track of properties essential
for estimating their \x\ luminosity, \lx,
as a function of time. These properties
include the mass-transfer rate, $\dot{M}$,
as well as the mass and radius of the accretor, $M_a$
and $R_a$, respectively.
We also identify BH or NS accretors, 
transient or persistent sources, 
evolutionary stages of donors and donor masses. 

\subsubsection{RLOF systems}
For persistent systems, \lx\ is estimated as

\exi L_X = {\rm min} \left( f L_{\rm Edd} , \eta_{\rm bol} \epsilon \frac{G M_a
  \dot{M} }{ R_a } \right) \ ,
\label{equ:Lx_per}
\exo 

\noindent with $f$ equal to unity. 
%
The value of $R_a$ is 10 km for a NS and 3
Schwarzschild radii for a BH, $\epsilon$ gives a conversion efficiency
of gravitational binding energy to radiation associated with accretion
onto a NS (surface accretion, $\epsilon = 1.0$) or BH (disk
accretion, $\epsilon = 0.5$), and $\eta_{\rm bol}$ is a factor that
converts the bolometric luminosity to \lxtt, the X-ray luminosity in the
full \chandra\ energy band, consistent with our observations. 
We use results from the literature
to tabulate the best available
estimates of this factor and its $1\sigma$ uncertainty for neutron
star and black hole accretors in \lq\lq low-hard\rq\rq\ and \lq\lq
high-soft\rq\rq\ state systems (see F13 for a detailed discussion
and references).
%
%
Since we are interested in combining our PS models with galaxy star
formation histories, we define lookback-time windows, $\delta t_{\rm lb}$,
in which we evaluate the total stellar mass produced in the galaxies
(see \scr{sec_massnorm}). In the
\stk\ simulations each model source is associated with its own
time-step window, $\delta t_{\rm step}$, which we use to calculate the fractional
time, $\delta N$,
that a source is on in a
given galaxy lookback time interval, i.e.

\exi
\delta N \equiv \delta t_{\rm step}/\delta t_{\rm lb} \ .
\label{equ:dN}
\exo

%
%

For transient systems we follow the prescription of F08,
calculating the outburst \x\ luminosity depending
on the accretor type, either BH or NS, via

\exi
L_{\rm X}=\eta_{\rm bol}\epsilon\times \Bigg\{
\begin{array}{l}
\min{\left(f_{\rm BH} \times L_{\rm Edd} , 
f_{\rm BH} \times L_{\rm Edd} \left[\frac{P}{10 \rm h}\right]\right)}, \rm{for\ BH} \\
\\
\min{\left(f_{\rm NS} \times L_{\rm Edd} , 
\frac{GM_a\dot{M}_{\rm crit}^{2}}{R_a\dot{M}_d} \right)}, \rm{for\ NS}\\
\end{array}
\label{equ:Lx_outb}
\exo

Here $f_{\rm BH}, f_{\rm NS}$ are factors setting an upper limit for the maximum
Eddington luminosity for black holes and neutron stars, equal to 2 and 1,
respectively. $\dot{M}_{\rm crit}$, $\dot{M}_d$ and $P$ are the
critical mass transfer rate for thermal disk instability,
the rate at which the donor star is losing mass and
the orbital period, respectively.
In addition, transient systems 
are mostly in a quiescent state and are too faint to be detectable, 
except when they go into outbursts. The fraction of time they are
in outburst defines their duty cycle, $DC$.
Following \citet{fragos2008},
we estimate $DC$
as 
\citep{dobrotka2006}
\exi
DC = \left( \frac {\dot{M_d} } { \dot{M}_{\rm crit} } \right) ^ 2 \ ,
\exo
for NSs and 5\%\ for BHs \citep{tanaka1996}.
For transient systems it then follows that
the fractional time a source is on in a
galaxy lookback time interval becomes
$\delta N \equiv \delta t_{\rm step}/\delta t_{\rm lb} \times DC$.
\vspace{1cm}

\subsubsection{Wind-fed systems}
For wind systems the \lx\ calculation is as for
for persistent RLOF systems (\er{equ:Lx_per}),
while the fractional lookback time is once
more given by \er{equ:dN}.

\subsection{Stellar Mass Normalization}\label{sec_massnorm}
The \stk\ simulations described thus far link
stellar mass described by a set of physical parameters with
an XRB population but do not contain information to link
the XRB population to host galaxy properties.
Thus the stellar mass produced by a given model is in a sense
arbitrary.
To construct tXLFs corresponding to real galaxies,
we need to modify this arbitrary mass by making use of
the galaxy's star formation history. This in turn modifies
the numbers of XRB sources produced by the model.
For the galaxies in our sample, SFH estimates
exist based on SED fitting by
\citet[][see \tr{tab-sfh}]{noll2009}. We use these results to calculate the total stellar
mass produced for each SINGS galaxy in our sample
in each lookback time window. Noll et al. have used
the SED fitting code CIGALE and assummed two exponentially
varying star formation rates for a young and an old 
population.
Using their symbols (\tr{tab-sfh}), the total SFR
at a lookback time $t_{\rm lb}$ is given by

\begin{eqnarray}
 {\rm SFR}_{t_{\rm lb }} &=&  \frac{  f_{\rm burst}   M_{\rm gal} }{ \tau_{\rm ySP} \left( e^{ t_{\rm ySP} / \tau_{\rm ySP} } -1 \right)} e^{ t_{\rm lb}/\tau_{\rm ySP} }  \nonumber \\ 
                           &+&  \frac{(1-f_{\rm burst}) M_{\rm gal} }{ \tau_{\rm oSP} \left( e^{ t_{\rm oSP} / \tau_{\rm oSP} } -1 \right)} e^{ t_{\rm lb}/\tau_{\rm oSP} }   \nonumber \\ 
\end{eqnarray}

As in \tr{tab-sfh}, \mgal\ is the total mass of stars and gas
that originates from stellar mass loss in
the galaxy, $f_{\rm burst}$ is the mass
fraction of the young population, $\tau_{\rm ySP}, \  \tau_{\rm oSP}$ are
the e-folding timescales for the young and old stellar population,
and $t_{\rm ySP}, \ t_{\rm oSP}$ the ages of the young and old stellar population.
All of these parameters are SED fit results
allowing us to calculate SFR$\bm{ _{ t_{\rm lb}}  }$ via equation 5,
and then stellar masses, used to scale XRB numbers for each galaxy-model pair.

\subsection{Construction of theoretical XLFs for SINGS galaxies}
As explained in F13, the bolometric corrections used to convert \stk-derived
\lx\ values for model sources to the full \chandra\ band
are empirical and introduce an uncertainty to the
estimated X-ray luminosity value.
Thus each \lxtt\ value can be thought of as
the mean of a Gaussian
distribution with a standard deviation originating in
the uncertainty introduced by the bolometric correction.

%
%
Further, as explained above, for each model source 
the fractional time a source is on
in a galaxy lookback time window is given by $\delta N$.
This number 
can also be
considered as the mean of a Poisson distribution that represents
the expected
number of times that this source will appear in the XLF. For each
source, we use this information to draw a random Poisson deviate,
giving a number of times that this source will be on. For each 
case that the source is on,
we draw a random Gaussian deviate from the \lx\ distribution
giving an \lx\ value for that
appearance. When this procedure is complete for all sources, we obtain
a set of \lx\ values, which can be converted to an XLF. However, this
is a single realization.  To obtain a reliable estimate for the mean
XLF and its uncertainty, we carry out 500 Monte Carlo realizations of
this process for the total XLF, and 100 for subpopulation
XLFs (donor/accretor type, LMXB/HMXB).
This procedure gives a reliable estimate for the mean number of
sources in each \lx\ bin, as well as $\pm 1\sigma$
uncertainties.

Overall, we thus obtain $12\times192=2304$ tXLFs (in both cumulative
and differential form) combining SFHs for 12 SINGS galaxies
and 192 \stk\ models. The tXLFs for the best models
(\scr{sec_lik}) are shown as red dotted curves
in \fr{fig_otxlf}.

\subsection{Background contamination}\label{sec_bg}
Our purpose is to compare our theoretically derived tXLFs for
XRB populations to oXLFs of point sources in
SINGS galaxies. 
Whereas all theoretically obtained point sources in the tXLFs
are known to be XRBs by construction, this is not
necessarily the case with all observed point sources
in SINGS galaxies.
It is possible that some of the latter
are actually background AGN rather than galactic XRBs.
We correct our pure XRB-based tXLFs by adding a 
component that takes into account the expected number
of background \x\ point sources in the area surveyed for
each galaxy. 

We correct our differential and cumulative theoretical
XLFs as follows.
We use the $\log N - \log S$ results of
\citet{kim2007} ($\Gamma=1.7$, broad band B,
their Table 3)
to estimate the number of
background sources expected in each luminosity
bin of our theoretical XLFs, taking into account the
area observed for each galaxy (either the \chandra\ S3/I0-I4 
detector area
or the $D_{\rm 25}$ region, whichever is smaller). We then
add these numbers to our \lq\lq pure XRB\rq\rq\
source numbers in each luminosity bin, thus obtaining
background corrected tXLFs. In \fr{fig_otxlf}
these corrected, final tXLFs are shown as solid 
red curves for highest-likelihood models
(\scr{sec_lik}). A comparison with observational completeness corrections
(see the blue curves in \fr{fig_otxlf}), which only affect
the faint end of the XLF, shows that
the background correction affects the XLF over the full
range of \x\ luminosities.

\section{Likelihood Functions}\label{sec_lik}
To obtain a quantitative estimate of the level of agreement
between observational and theoretical XLFs,
we construct and evaluate likelihood
functions, using the differential XLF versions.
As explained below, we calculate two types of
likelihoods.

\subsection{Pair likelihoods for a given galaxy}
For each galaxy there is a unique, completeness corrected oXLF coming from
our \chandra\ observations (continuous blue curves in \fr{fig_otxlf}). 
We wish to establish
which of the 192 \stk\ models used in this work,
after it has been combined with a specific galaxy's SFH
and corrected for background AGN,
best describes this oXLF. In other words, there
are 192 tXLFs for this galaxy that need to be
compared with a single oXLF. Thus, we 
calculate pair likelihoods for 192 o-t XLF pairs.

Given a galaxy and its oXLF, $k$, and a model tXLF, $m$,
we define the {\it $k-m^{th}$ pair likelihood},
\lpairkm, as the probability, $P$, of obtaining an
observational set of XLF data points, $k$, given a theoretical
model XLF, $m$. This is given by 
\exi
{\cal L}_{{\rm pair,} km} = \prod_i P_{\rm Poisson} (N_{{\rm obs},i}, N_{{\rm th},i} )  
\exo
\label{equ:lik_th}

Here $P_{\rm Poisson} (N_{{\rm obs},i}, N_{{\rm th},i})$ is the
Poisson probability of observing $N_{{\rm obs},i}$ point sources, 
in the $i^{\rm th}$ luminosity bin,
treating
the theoretically obtained number of point sources, $N_{{\rm th},i}$, in the
same bin, as the expectation value for
the number of sources at this luminosity. 
The total pair likelihood is the product of
all such probabilities over all luminosity bins. This compares
the agreement of the two XLFs in each luminosity bin, and
thus its overall shape and normalization.

Note that tXLFs are calculated for 100 equal-sized bins in
log~$   L_X  $, spanning the observational range of luminosity values.
The oXLFs are binned to match the tXLFs bins. In addition,
in order to compare corresponding quantities,
$N_{{\rm obs},i}  $ are the observed
numbers {\it before} correction for incompleteness, while
$  N_{{\rm th},i}  $ are the theoretical numbers after
the observationally derived incompleteness information for a given bin 
has been taken into account.

For a given galaxy, the best model is that for which this
procedure produces the highest pair likelihood value. 
We tabulate highest-likelihood results based on this procedure in
\tr{tab-bestgalmod} for all galaxies in our sample.  Since the
absolute numeric value of \lpairkm\ obtained via \er{equ:lik_th} has
no particular meaning, in column 10 each \lpairkm\ value is normalized
to the highest value in the table. This leads to a ranking, shown in
column 9, with the highest \lpairkm\ value having rank 1.  We also
plot likelihood results against model number for all o-t pairs (i.e. not just highest-likelihood
pairs) in the top panel of \fr{fig_lik_pair_glob_mods}. 
For convenience, in this figure \lpairkm\ values are
normalized so that they range from 0 to 1 (see \scr{sec_lnorm}).

\subsection{Global likelihoods}
The previous procedure determines
the best model for the $k^{th}$ oXLF by estimating 
the probability, \lpairkm, of obtaining this XLF with
each model. 
We are also interested in knowing the best model, $m$, for {\it all}
oXLFs taken as a whole.  To determine this, we calculate the product of all
\lpairkm\ values by defining the {\it global likelihood}

\exi
{\cal L}_{{\rm global}, m} \equiv \prod_k {\cal L}_{{\rm pair}, km}
\label{equ:glik_th}
\exo
where \lpairkm\ is given by \er{equ:lik_th} and 
the index $k$ runs from 1 to 12, corresponding
to each of the 12 oXLFs. The results of this procedure are tabulated
in \tr{tab-bestbymod15}, ranked by the 
ratio of each global likelihood value to
the maximum likelihood value in the table
(model 245). In \fr{fig_otxlf} we also show
for each galaxy the tXLF that corresponds to
the best global model (245) as a dark grey curve.

\subsection{Likelihood Normalization}\label{sec_lnorm}
The specific pair or global numeric likelihood values obtained
have no particular meaning, except relative to each other.
For plotting purposes (Figures 
\ref{fig_lik_pair_glob_mods}, \ref{fig_likparam})
it is convenient to
normalize likelihood values
so that the maximum 
value is 1 and the minimum 0. 
This is done as follows.

Let $\cal L$ stand for either \lpair\ or \lglob.
Then the final normalized likelihood value is given by
\exi
\Lambda_{\rm n} \equiv \Lambda_{\rm min} / {\rm max} (\Lambda_{\rm min}) \,         
\exo
where
\exi
\Lambda_{\rm min} \equiv \ln [ {\cal L} / {\rm min} ({\cal L}) ] \ .              
\exo 
In what follows we use an additional subscipt to specify whether
the final normalized likelihood is a pair or a global likelihood
($\Lambda_{\rm n, pair}   $ and $   \Lambda_{\rm n, global}   $, respectively).

\section{Results and discussion}
\subsection{Best models}\label{sec_best}

A visual inspection of o-t XLF pairs in \fr{fig_otxlf} suggests that
in many cases tXLFs (solid red curves) are successfully
reproducing oXLFs (solid blue curves) both in shape and
normalization. However, there is considerable variation and room for
improvement. The best case is NGC 4826 and Model 269, which based on
its likelihood estimate has rank 1 in \tr{tab-bestgalmod}. The tXLF
for this galaxy and Model 245 (our best global model), shown by the
dark 
grey line in the panel, is also very close to the oXLF.  The worst case is NGC
3184 and Model 277 (rank 12 in \tr{tab-bestgalmod}).  Since likelihood
estimates take into account source numbers in luminosity bins
(\er{equ:lik_th}), the low likelihood estimate in this case is driven
by the strong discrepancy in numbers at low luminosities.  In many
cases XLFs for the best pair model (red curve) and best global model (dark grey
curve) are very close, but there are also cases where these are
in strong disagreement.  

The top panel of \fr{fig_lik_pair_glob_mods} displays the normalized pair likelihood, \lapairn, for
all $192\times 12$ o-t pairs (all 12 galaxies), versus all models used
in this paper. The imposed normalization (see \scr{sec_lnorm}) is such
that the maximum normalized likelihood for each o-t pair is 1 and the
minimum is 0.  Curves plotting normalized likelihood values as a
function of model number (1-96 and 193-288) for each galaxy/o-t pair
are shown with a different color.  It is immediately obvious that for
each galaxy there are a number of good models, indicated by maxima in
the curves, as well as a number of poor models indicated by minima.
Closer inspection of the plot reveals the remarkable fact that,
overall, there is strong clustering of local maxima and minima,
indicating that good models are good for {\it all} galaxies (and vice
versa for poor models).  This is so despite the fact that for a given
galaxy the best models do not always match oXLFs well
(\fr{fig_otxlf}).  This suggests that, {\it to first order}, the
parameters associated with a good model are useful indicators of XRB
physics across all 12 galaxies.

This global trend is corroborated by the lower panel
of \fr{fig_lik_pair_glob_mods} which shows
the normalized global likelihood, \laglobn. As above, the normalization
is such that the maximum is 1 and the minimum 0. Given the similarity
of \lapairn\ curves (top panel) with each other, it is not
surprising that they are also similar with the global likelihood
curve, \laglobn, as the latter combines, for a given model, all pair
likelihoods for all galaxies (\er{equ:glik_th}).

To understand what highest-likelihood results imply for
individual model parameters, we tabulate results in two ways.  
We
first rank 
highest-likelihood galaxy-model pairs according to {\it pair}
likelihood value and show the results in \tr{tab-bestgalmod}. 
Clearly, this
tabulation favors a \acol\ value of 0.1, a high-end IMF exponent
of $-2.7$ and a mixed initial $q$ distribution.
Second, we rank
all 192 models according to {\it global} likelihood value
and show the results for the 15 highest ranked models in
\tr{tab-bestbymod15}\footnote{The full table for all 192 models is
  available online.}.  These correspond to the 15 highest peaks in
the lower panel of \fr{fig_lik_pair_glob_mods}.  These results also favor 
a
\acol\ value of 0.1, and to a lesser extent a high-end IMF exponent of
$-2.7$ and a mixed initial $q$ distribution. 

By comparing Tables \ref{tab-bestgalmod} (col.~2) and
\ref{tab-bestbymod15} (col.~1), we note that all six best global models
are also among the best pair models.  This just underlies the fact
that these are the best models overall.  Conversely, nine of the best
pair models are also among the best global models.  For the three
galaxies NGC 1291, NGC 2841 and NGC 4736, the best pair models are not
among the 15 best global models.  It is, however, remarkable that
for nine out of twelve galaxies one of the best 15 models that
represent global averages over all galaxies also describes the
individual galaxy XLF.

We note that for galaxy NGC 1291 \citet{luo2012} use
\chandra\ observations that are deeper than ours to perform a detailed
study of the point source population.  They show that high-luminosity
($> 5\times 10^{38}$\lunits) LMXBs in NGC 1291 are likely associated
with a younger stellar population in the galaxy's ring. Their deeper
observations allow them to separate the bulge from the ring
population.  Even so, we note that our results still support their
conclusions, as we find strong LMXB contributions both from our old
and young populations.

\subsection{Deviations between models and observations}\label{sec_dev}
A key reason for deviations of oXLFs from tXLFs
is the presence of two \lq\lq jumps\rq\rq\ at two
different luminosities ($\sim 10^{37}$\lunits\ and a few $\times
10^{38}$ \lunits), which are not expected from observations.  There
are two reasons for this.  First, in our \stk\ models we identify
transient and persistent sources by comparing the calculated
mass-transfer rate of the binary to a critical mass-transfer rate,
below which the thermal instability develops, giving rise to transient
behavior.  Furthermore, we strictly limit the accretion rate to the
Eddington limit. These limiting mass-transfer rates in our modeling
are responsible for the jumps in tXLFs. However, in nature there is no
sharp transition between thermally stable and unstable disks, nor a
precise limit to the highest accretion rate possible. Accretion onto a
compact object is a non-linear and much more complex process, which in
reality can result in a smoother luminosity distribution with no sharp
transitions.  Second, the assumption of solar metallicity in practice
imposes a maximum BH mass of $\sim 15$~\msun.  If for instance part of
the stellar population had a lower metallicity (e.g. 30\% solar) then
the maximum BH mass would increase to $30$~\msun\ or more
\citep{belczynski2010}, which could smooth out the jump at the very
luminous end of the XLF.  

More generally, it must be stressed that the SFHs we use are very
simple.  \citet{noll2009} note that, as SED fitting is computationally
intensive, they select a limited set of values for their SED parameter
model grid.  Thus the age of the old population is a constant (10 Gy),
and there are only two possible values for the age of the young
population, 200 Myr and 50 Myr.  The parameter that was allowed to
vary the most is \lq\lq $f_{\rm burst}$\rq\rq, the mass fraction of
the young stellar population at the present time (nine possible
values).  Further, their detailed analysis of their SED fitting
results shows that best-fit e-folding times and the age of the young
population are highly uncertain, due in turn to photometric errors and
uncertainties in stellar population models.  Finally, as already
mentioned, their SFHs have uniformly solar metallicities. Although
this is based on the best estimates to date, note that these are based
on gas metallicities with their own set of significant uncertainties
\citep[see][where two sets of metallicity results are presented,
  straddling solar metallicity for all galaxies in this
  sample]{moustakas2010}.  Further progress in matching oXLFs will
require more detailed SED fitting.

\subsection{Constraining Parameters}
\tr{tab-bestgalmod} and especially \tr{tab-bestbymod15} suggest likely
best values for model parameters in our model grid. 
As mentioned, both tables suggest \acol~$\simeq 0.1$.
In
\fr{fig_likparam} we investigate this further by plotting resistant
mean\footnote{http://idlastro.gsfc.nasa.gov/ftp/pro/robust/\newline
  resistant\_mean.pro} values for the normalized global value,
\laglobn, vs. the four \stk\ \acol\ parameter values used in our
simulations.  The calculation of the resistant mean iteratively
rejects outliers beyond $3\sigma$. This is useful for highlighting the
clustering of likelihood values, which often show a considerable
spread.

\fr{fig_likparam} shows that higher likelihood values are
systematically favored for models with \acol~$\simeq 0.1$:
Not only do \laglobn\ values converge to their highest value as \acol~$\rightarrow
0.1$, but they also do so with a decreasing spread, as indicated by the $3\sigma$
resistant mean error bars.

\tr{tab-bestgalmod} and \tr{tab-bestbymod15} also appear to favor a
value of $-2.7$ for the high-end slope of the IMF.
Although a value of $-2.7$ agrees with some estimates
\citep[e.g.][]{scalo1986}, we note that it is somewhat steep compared
to the so-called \lq\lq canonical\rq\rq\ value of $-2.3$ established
observationally for resolved stellar populations in the local group
\citep{bastian2010,kroupa2012}.  In spite of the fact that (1)
variations with metallicity and star formation rate density for the
integrated galaxy IMF remain hotly debated \citep[][ and references
  therein]{kroupa2012} and (2) taking uncertainties in the canonical
value into account, a value of $-2.7$ is still close to the canonical
upper limit, we do not consider a value of $-2.7$ to be a robust
result.  This is because, first, the Noll et al. SED fitting results
are {\it assuming} the canonical value, and it is these results that have been
convolved with our \stk\ models. Second, four out of the fifteen best
global models do, in fact, agree with the canonical value.

For the remaining parameters these tables show that the results are
inconclusive. In addition, for most of these parameters we only use
two different values in the \stk\ simulations, so we cannot investigate
any trends such as those for \acol\ in \fr{fig_likparam} .

Given the caveats for the IMF slope, the results for
\acol~$\simeq 0.1$, and to some extent for the prevalence of a mixed
initial $q$ distribution provide the most significant constraints for
binary star parameters from this work. 

As \acol\ is a combination of two parameters, we are unable to set any
constraints on either $\lambda$ or \aco\ individually.  In their
investigation of Galactic merger rates for compact objects,
\citet{dominik2012} set \aco~$=1.0$ and study the behavior of
$\lambda$. The latter is not considered constant throughout the
evolution of the donor, but depends on donor parameters such as mass,
radius and evolutionary stage.  They find $\lambda$ values between
$\sim 0.1$ and $\sim 0.2$ for NS progenitors, and below $\sim 0.1$ for
BH progenitors. With the assumption of \aco~$=1.0$ our result for
\acol\ would thus be largely consistent with the detailed stellar
evolution models of \citet{dominik2012}.

\subsection{Comparison with F13 and T13}\label{sec_comp}
F13 and T13 use the same \stk\ models as this paper but combine them
with galaxy information from the Millennium II simulation and the
semi-analytic galaxy catalog of \citet{guo2011}.  While F13
investigate the total galaxy specific \x\ luminosity (\lx/SFR,
\lx/\mstar) evolution out to $z\sim 20$, T13 construct galaxy XLFs for
comparison with observations out to $z\sim 1.4$ and make predictions
out to $z\sim 20$. Thus our three papers examine XRB formation and
evolution in different contexts, and also use different likelihood
formulations. To investigate whether there is agreement in
highest-likelihood models between the three papers, in
\tr{tab-bestbymod15} we also show ranks from F13 and T13 for our best
15 global likelihood models.  There is good agreement overall, and in
certain cases the agreement is exceptionally good.  In particular, our best
ranked model (245) is also F13's reference model (their rank 1) and
has rank 4 in T13.  Models 245, 277, 229, 205, 269, 249, 273, 37 and
85 are all within the 15 highest ranked in all three papers.  This is
further evidence in support of \acol~$\simeq 0.1$ both in the local
universe and over cosmic time.

\subsection{XLF subpopulations}\label{sec_subpop}
In Figures~\ref{fig_mass_oy} and \ref{fig_da} we plot the total tXLF
for each galaxy and 
highest-likelihood model (black continuous curve), together with
constituent subpopulation tXLFs.

We define two sets of subpopulations.  The first
is shown in \fr{fig_mass_oy} where we plot
XLFs for LMXBs (red) and HMXBs (blue).  In our
simulations XRBs are labelled HMXBs if the donor star has mass
$M_{\rm donor} \ge 3M_{\odot}$, and LMXBs otherwise.  The mixture of
populations present in each galaxy is closely tied to the assumed SFH
(\tr{tab-sfh}).  To illustrate this, XRBs originating in the \lq\lq
old\rq\rq\ stellar population are shown by the dashed curve in
\fr{fig_mass_oy}, while those originating in the
\lq\lq young\rq\rq\ SP are shown by the solid
curve. Note however that
the terms old and young can be misleading
in this context. Although the two are distinct in terms of age, with
the young one appearing several Gyr after the old one, it is possible
for the latter to still be actively star-forming, depending on the
e-folding timescale $\tau_{\rm oSP}$. In other words, an
\pold\ population is not necessarily a \lq\lq red and dead\rq\rq\ one.

The trends seen in \fr{fig_mass_oy} for each galaxy can be understood
qualitatively by referring to (1) the SFH (\tr{tab-sfh}), which we
combine with \stk\ models to construct tXLFs,
and (2) \tr{tab-bestgalmod}
which gives the details of each best model for each galaxy.  
We should
also keep in mind the relative evolutionary timescales for LMXBs and
HMXBs. As shown in F13\footnote{F13 use the same definition for LMXBs
  and HMXBs as this paper in terms of a threshold donor mass of
  3\msun.}  (their fig.~2), in a single burst population
the contribution in \x\ output from HMXBs
peaks at about 5 Myr, remaining important up to an age of $\sim
100-300$ Myr, while LMXBs take over at $\sim 100-200$ Myr.  In terms
of the assumed \pyoung\ and \pold\ SPs, a population's contribution
will be more significant as its e-folding timescale is longer and age
is younger.  A higher young mass fraction will tend to
increase the contribution from the young population.

Looking at galaxies NGC 1291 and 2841, we note that
they have all best fit SFH parameters the
same, and slightly different \mgal, 
while the associated \stk\ model is the same.
It is evident that the subpopulation XLFs are similar, with
the old LMXB SP dominating in both cases.

For galaxies NGC 3184 and 3627, the main SFH difference is a higher
$\tau_{\rm oSP}$ for the latter. Based on this, one might expect that
in NCG 3627, the old population would be more dominant. However, young
SP HMXBs dominate for most of the XLF. This is due to \stk\ model 245
(NGC 3627) having half the stellar wind strength compared to model 277
(NGC 3184). Weaker stellar winds lead to smaller mass loss for
primaries that eventually become compact objects, and thus to numerous
and more massive BH XRBs. The latter in turn tend to be more luminous
than NS XRBs as (1) they can form stable RLO XRBs with massive
companions, and (2) they show higher accretion rates due to the high BH
masses.  Although a weaker stellar wind also decreases the accretion
rate in wind-fed HMXBs, it turns out that this is not the dominant
effect (F13).

NGC 3198 has by far the highest $\tau_{\rm oSP}$ among all galaxies
and this is reflected 
in the dominance of the old LMXB population.
In contrast, NGC 3521 and 5055 have a much lower
$\tau_{\rm oSP}$ (with other parameters except for \mgal\ identical
for all three) and are dominated by young SPs. NGC 4631 is also
similar, with a somewhat lesser contribution from the old SP
(higher $\tau_{\rm ySP}$). Compared to NGC 4631, NGC 4826 has lower
$\tau_{\rm ySP}$ and $f_{\rm burst}$ leading to a relatively
stronger contribution from the old SP.

NGC 4736's young SP has the smallest best-fit age (50 My) together with
a high $\tau_{\rm ySP}$. This is consistent with the clear
dominance of young HMXBs, which are closest to their peak activity
timescale \citep{shtykovskiy2007}, 
while LMXBs haven't yet had time to contribute
significantly to the total \x\ luminosity. 

Finally, NGC 5474 has higher $\tau_{\rm oSP}$ but a higher young mass
fraction compared to, e.g. NGC 3521, so it is moderately dominated by
the young SP.

The second subpopulation set is shown in \fr{fig_da} in
terms of donor and accretor stellar type.  We classify donors as
MS, evolved or degenerate.  Evolved types have left the MS and can be
in any of a number of different stellar evolutionary stages, including
the Hertzsprung gap, the red giant branch (GB), core helium burning,
early asymptotic GB, thermally pulsating asymptotic GB, helium MS,
helium HG or the helium GB. Degenerate types are helium or
carbon/oxygen white dwarfs.  On the other hand, accretors are either
BH or NS.

A number of trends are visible in these plots.  Considering the
accretor subpopulations, indicated by different linestyles in
\fr{fig_da}, we notice that in general BH accretors (solid lines)
dominate the XLFs at high luminosities.  In addition, galaxies with no
sources at \lx~$\gtrsim$~\ten{3}{38} have no BH accretors.  These
galaxies are NGC 3184, 3351, 3521, 4631 and 5055.  The reason for this
is that, since BH accretors originate in the high-mass end of the IMF,
from a statistical point of view on average there will be fewer such
systems. In addition, BH XRBs can have higher luminosities than NS
XRBs, but not vice versa. Thus, BH XRBs can only be registered only
when very luminous sources are present \citep[e.g. see][for NGC
  4649]{luo2013}, as NS XRBs will always dominate at low luminosities.
As a result, for galaxies with very few or no sources above $\sim
10^{38}$\lunits\ models are unable to constrain the BH XRB population.

All donor subpopulation XLFs, indicated by different colors in
\fr{fig_da}, are dominated by MS donors at low luminosities ($<
10^{37}$\lunits)\footnote{This is not always obvious in \fr{fig_da} as
  the \lx\ ranges shown are aimed to match the observed XLF ranges.}
and by evolved donors at high luminosities. This is
a consequence of the fact that evolved donors are mainly
giants in RLOF systems, so they all have large accretion disks and
long orbital periods. Since
$\dot{M}_{\rm crit}$ 
depends on the size of the accretion
disk and orbital period, it follows that in the case of
these systems it will always be higher than
a threshold value, which corresponds to \lx~$\sim 10^{37}$\lunits\
\citep{king1996,dubus1999}. Further, systems with
mass transfer rates less than $\dot{M}_{\rm crit}$ and, thus,
with luminosities lower than $\sim 10^{37}$\lunits\ will be transients,
which are mostly quiescent. Thus in practice at \lx~$\gtrsim 10^{37}$
the dominant donor systems will be persistent systems
with evolved donors.

\subsection{Comparison with Lehmer et al. 2010}\label{sec_L10}
It is well established that emission from HMXBs and LMXBs correlates
with galaxy-wide SFR and \mstar, respectively. This is due to the fact
that the former are relatively young ($\lesssim 100$ Myr) compared to
the latter ($\gtrsim 1$ Gyr).  \citet[][L10]{lehmer2010} parametrize
the contribution of LMXBs and HMXBs to the $2-10$ keV integrated
luminosity in star-forming galaxies as 
\exi L_{\rm HX} = L_{\rm
  HX}^{\rm LMXB} + L_{\rm HX}^{\rm HMXB} \equiv \alpha \ M_{\star} +
\beta \ {\rm SFR} \ ,
\label{equ:L10}
\exo 
where \lhx~$\equiv L_{X,{\rm 2.0-10.0 keV}}$.  They use a sample
of galaxies spanning a large range in star-forming activity observed
with \chandra\ to obtain best-fit values $\alpha = (9.05\pm0.37)
\times 10^{28} {\rm erg \ s}^{-1} M_{\odot}^{-1}$ and $\beta =
(1.62\pm0.22) \times 10^{39} {\rm erg \ s}^{-1} (M_{\odot} {\rm
  yr}^{-1})^{-1}$. We use these results to compare with our work,
noting also that they are consistent with other work both
on the \lx$-$\mstar\ relation in elliptical galaxies
\citep{gilfanov2004a,zhang2011,boroson2011} and on the
\lx$-$ SFR relation in galaxies with high specific SFR, where
HMXBs are dominant \citep{gilfanov2004b,mineo2012}.

We use our subpopulation XLFs for LMXBs and HMXBs to compare with the
L10 results as follows. We obtain galaxy-wide \x\ luminosities by
integrating each model XLF over all luminosity bins.  Note that for
each galaxy we use both the highest pair-likelihood model XLF and the
XLF which is based on the best global model 245, thus obtaining two
sets of galaxy-wide \x\ luminosities.  
We then use the \mstar\ and SFR
values for each galaxy to estimate $L_{\rm HX}^{\rm LMXB}$ and $L_{\rm
  HX}^{\rm HMXB}$ from \er{equ:L10}. Since our data are in the $0.3 -
10$ keV band, we convert \lhx\ values to \lxtt\ by using the mean
\lxtt\ to \lhx\ ratio based on the F13 models.  In \fr{fig_L10} we
plot the integrated \x\ luminosities for LMXBs and HMXBs against
\mstar\ and SFR. The grey open circles are for
integrated luminosities that use the highest
pair-likelihood model XLF, while the filled
black circles are for model XLFs using the best global model 245.
We also show the relations $L_{\rm HX}^{\rm LMXB} -
$~\mstar\ and $L_{\rm HX}^{\rm HMXB} - {\rm SFR}$ for the best-fit
$\alpha,\beta$ values of L10, with the associated $1\sigma$ scatter.

The comparison shows that, when the highest pair-likelihood models are
used, there is some agreement, especially for LMXBs, but some of these
these fail to reproduce the L10 relations.  In contrast,
when the best global model is used, integrated luminosities are
in 
much better agreement
with the L10 relations, at least for HMXBs. The residual scatter is
likely due to limitations of SFHs and uniformly solar
metallicities. The observed agreement suggests that
averaging over all twelve galaxies to calculate global likelihoods
produces results that, in a statistical sense, are more reliable. On
the other hand, individual pair-likelihoods inevitably suffer from all
SFH uncertainties discussed earlier, even though a given tXLF might
match a given galaxy oXLF better than the tXLF for model 245. This
result is fully consistent with the fact that model 245 is also the
top-ranked model of F13.  Using an independent likelihood formulation,
these authors constrain their best models by comparing with
observational work, including the L10 relations.

This result provides strong motivation for future work.
One would expect that for a larger dataset, such
as all 75 SINGS galaxies, agreement would be further improved.

\section{Conclusions}\label{sec_conc}
We have constructed theoretical XRB XLFs, for the first time corrected
for background contamination (\scr{sec_bg})
and including $1\sigma$ uncertainties,
for 12 nearby, late-type SINGS galaxies. We compare them to
observational XRB XLFs, corrected for incompleteness, by means of a
likelihood approach (\scr{sec_lik}).

Our main results are as follows:

\begin{enumerate}
\item By comparing 192 theoretical models (\scr{sec_stk}) to observed
  XRB populations in twelve nearby galaxies (\scr{sec_obs}), we are
  able to constrain best values for XRB formation and evolution
  parameters (Tables~\ref{tab-bestgalmod}
  and~\ref{tab-bestbymod15}). This is the largest scale comparison in
  terms of numbers of nearby galaxies and theoretical models to date.
\item There is substantial range in the level of agreement between
  observational and theoretical XLFs for individual galaxies, due to
  SFH and some model limitations (see \scr{sec_dev} and below).
For about half of the galaxies the agreement is not good.
  However, for any given model,
  likelihoods are consistently high or low both when estimated for
  individual galaxies (\lpair) and when averaged over the full galaxy
  dataset (\lglob). Thus parameters associated with highest-likelihood
  models provide insight for XRB physics irrespective of the details
  of specific galaxies (Figures~\ref{fig_lik_pair_glob_mods}
and \ref{fig_likparam}). 
\item Our best models have \acol~$\simeq 0.1$ and a mixed initial $q$
  distribution.  Their IMFs have high-end slopes
  of $-2.35$ or $-2.7$ and \ewind~$\simeq 1.0-2.0$ (Tables~\ref{tab-bestgalmod}
and \ref{tab-bestbymod15}, \fr{fig_likparam}). However,
we stress that 
further work is required before reliable values for these parameters can be
established (see below).
\item Our best models for XRBs in nearby galaxies are in agreement
  with work describing the cosmological evolution of XRBs (F13) as
  well as integrated XRB emission from entire galaxies (T13, see 
  \scr{sec_comp} and \tr{tab-bestbymod15}).
\item Model XLFs show considerable variation in their constituent
  systems.  Some galaxies have no BH accretors and most have a
  substantial contribution from LMXBs (\scr{sec_subpop}).
\item The integrated model XLF \x\ luminosity due to LMXBs and HMXBs
that is based on the best global model in this paper and in F13
agrees with the expectations from the L10 relations
  based on galaxy-wide stellar masses and, especially, SFRs (\scr{sec_L10} and
\fr{fig_L10}).
\end{enumerate}

This paper represents the first conserted effort to model
observational XRB XLFs for a set of late-type galaxies of this size.
For individual galaxies, tXLFs match oXLFs with varying degrees
of success. Even so, there are clear global trends  
regardless of individual galaxies.
We can thus begin drawing conclusions for a
number of physical parameters related to XRB formation and
evolution. 

The major limitations for this work come from the SFH, the small
sample, imposed limiting mass-transfer rates (\scr{sec_dev}) and poor
understanding of many aspects of physics related to XRBs, precluding
the construction of good models.  Observationally, this provides
motivation for increasing the sample to include more nearby galaxies
with reliable SFHs.  On the computational side, work on the physics of
XRB formation and evolution needs to include detailed modeling of the
common envelope phase, as well as detailed self-consistent mass
transfer calculations.  Poor or non-existent understanding of physics
remains a challenge for models. Thus, we do not really know how either
BH or Be XRBs form, although the latter constitute an important
population which forms the majority in the Small and Large Magellanic
Clouds.  We also do not yet understand the physics of disk
instability.

\chandra's superb angular resolution is critical for this type of work. Deeper
\chandra\ observations would allow to expand the dynamic range for comparisons
with models to fainter
luminosities, mitigating the need for completeness corrections. 

Deep $HST$ observations to securely identify counterparts
can also further our understanding of the nature of XRB systems.
Such
a task
has been successfully achieved to date only for a handful of nearby galaxies 
(at distances $\lesssim 10$~Mpc), uncrowded regions and the brightest
stars \citep[e.g.][]{dalcanton2009,rejkuba2009,tikhonov2012}.
In the medium term, the advent of 30-meter class telescopes such as the ELT and
TMT,
promises to pave the way to major breakthroughs in this field \citep{greggio2012}.

\acknowledgments 
We thank Stephan Noll for providing star formation
histories from SED fitting with CIGALE for the galaxies in this
sample.  
We thank Chris Belczynski for making \stk\ available to us.
PT acknowledges support through a NASA Postdoctoral Program
Fellowship at NASA Goddard Space Flight Center, administered by Oak
Ridge Associated Universities through a contract with NASA. TF is a
CfA and ITC prize fellow. BDL thanks the Einstein Fellowship
Program. KB acknowledges support from MSHE grant N203404939.  AH, PT,
AZ and VK were also supported by NASA ADAP 09-ADP09-0071
(P.I. Hornschemeier).  Computational resources supporting this work
were provided by the NASA High-End Computing (HEC) Program through the
NASA Center for Climate Simulation (NCCS) at Goddard Space Flight
Center and by the Northwestern University Quest High Performance
Computing (HPC) cluster.


\bibliographystyle{apj}
\bibliography{xlfb}    


\begin{figure*}
\epsscale{1.1}
\plotone{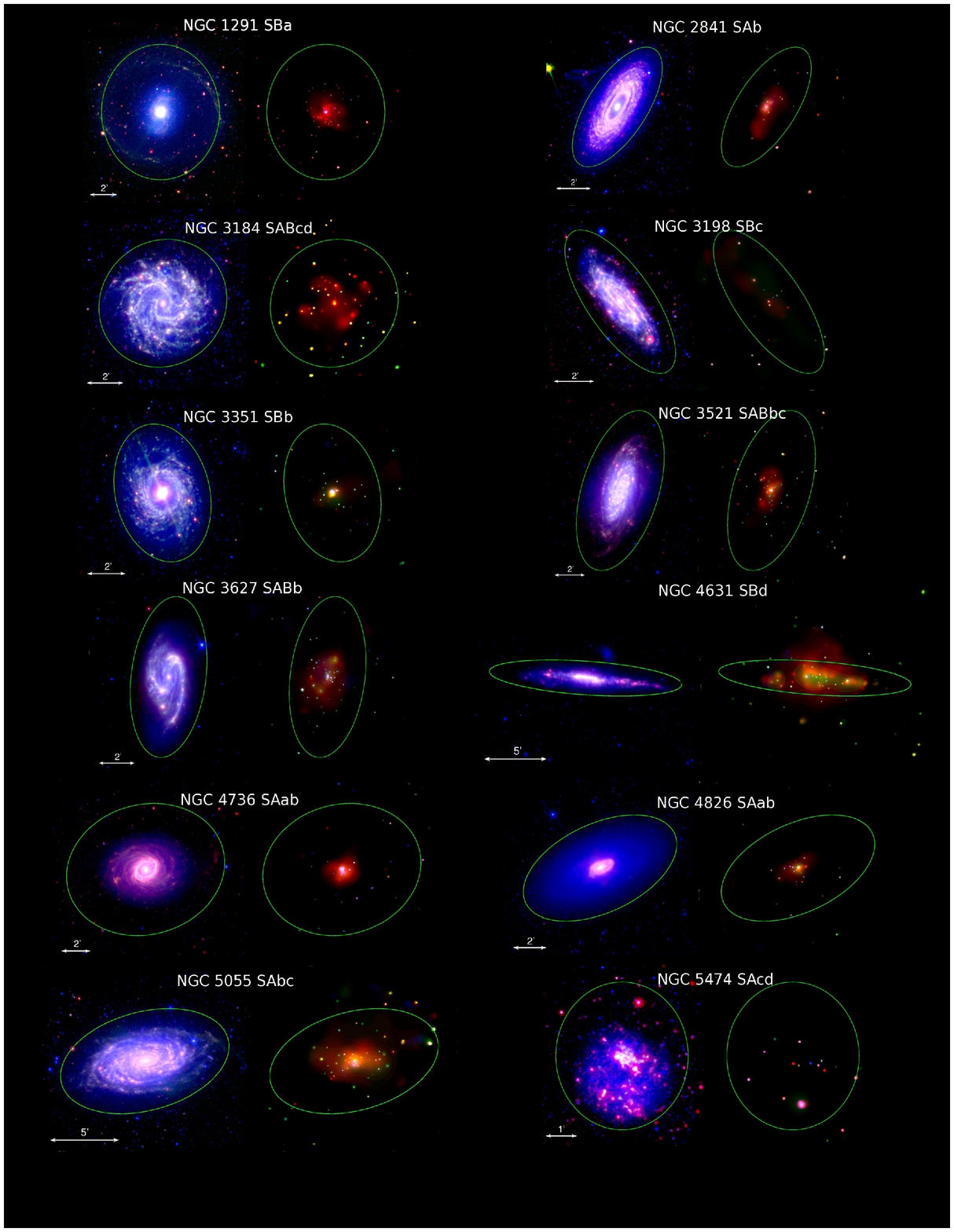}
\caption{Infrared morphologies and \x\ point sources of SINGS galaxies in this paper. Each thumbnail pair shows a \spitzer-IRAC composite false-color image at left (blue: 3.6$\mu m$, green: 4.5$\mu m$, red: 8.0$\mu m$) and a \chandra\ adaptively smoothed false-color composite at right (blue: 0.3-1.0 keV, green: 1.0-2.0 keV, red: 2.0-10.0 keV). The green ellipses indicate the $D_{25}$ isophotes. 
See 
\scr{sec_obs} for a brief description of \x\ point-source selection.
\label{fig_sample}}
\end{figure*}


\begin{figure*}
\epsscale{1.0}
\plottwo{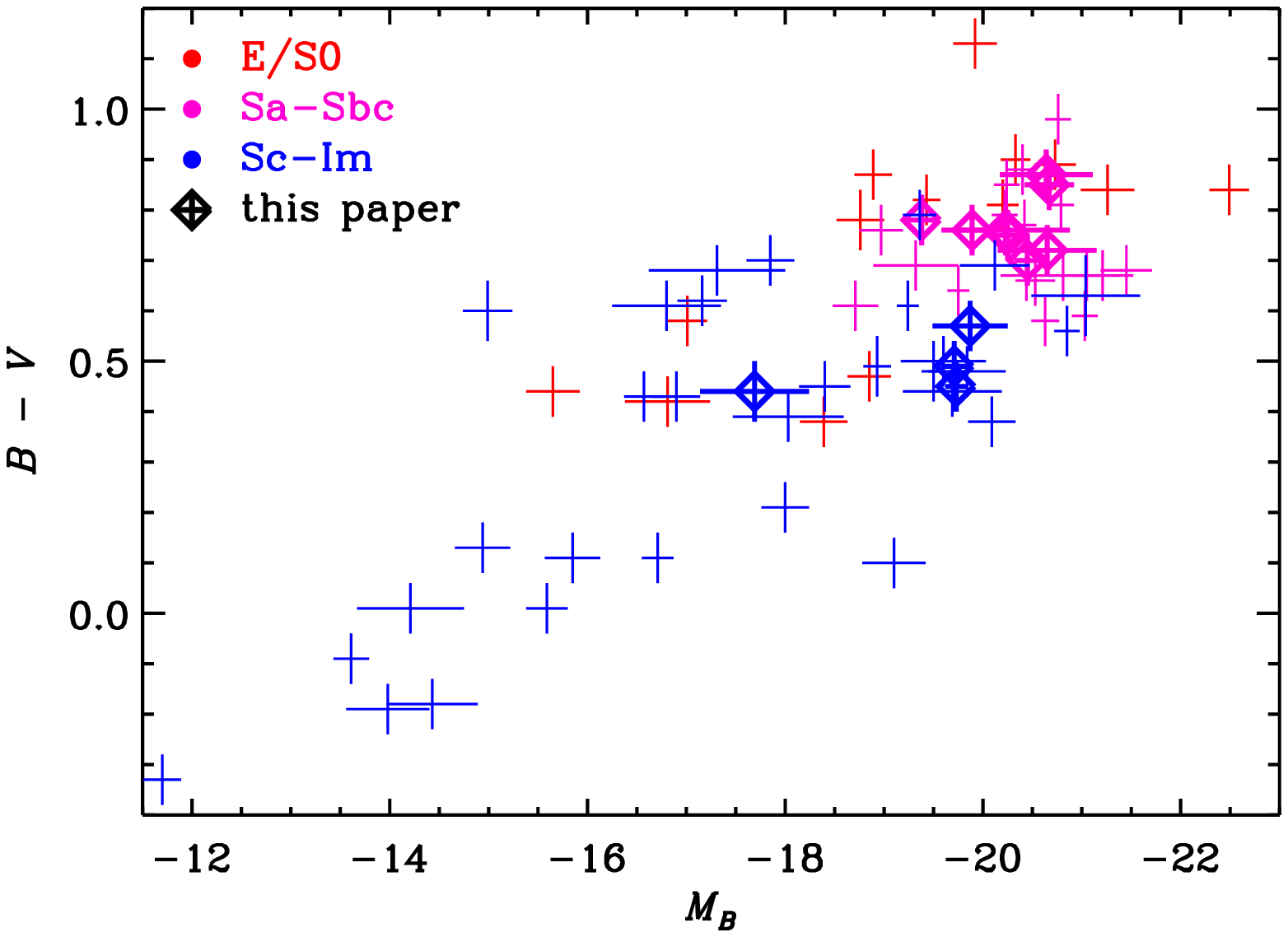}{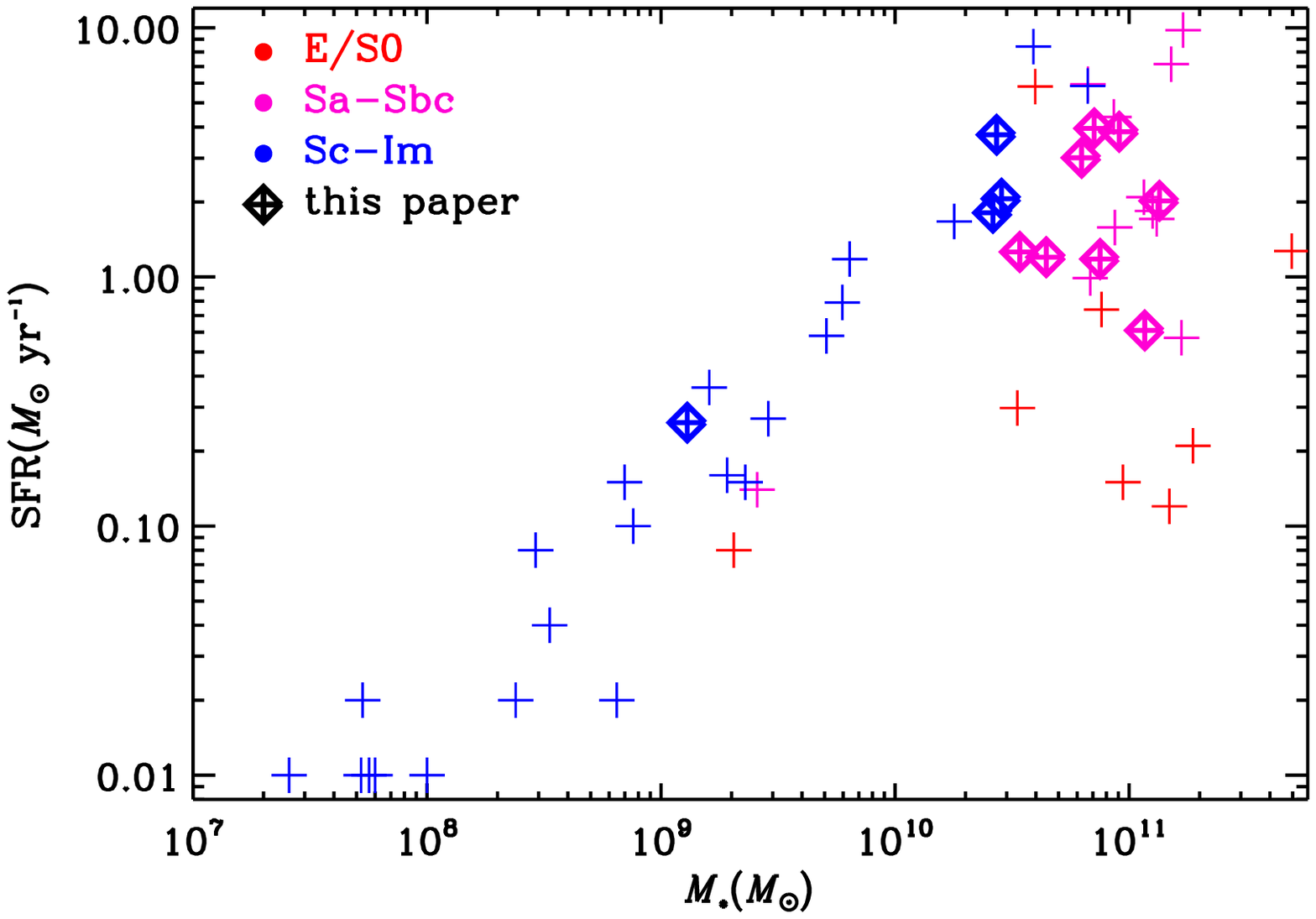}
\caption{Properties of our galaxy sample compared to other SINGS galaxies. 
{\it Left:} Color-magnitude diagram.
{\it Right:} Plot of SFR vs. \mstar. 
In both panels, crosses in normal font indicate 
SINGS galaxies that are not used in this paper. Bold crosses with diamonds
indicate the twelve SINGS galaxies used in this paper. Galaxies are color-coded
according to morphology as indicated in the legend.
Values in the color-magnitude plot are taken from \citet{moustakas2010},
and in the SFR vs. \mstar\ plot from \citet{noll2009}.
\label{fig_cmsm}}
\end{figure*}

\begin{figure*}
\epsscale{1.1}
\plotone{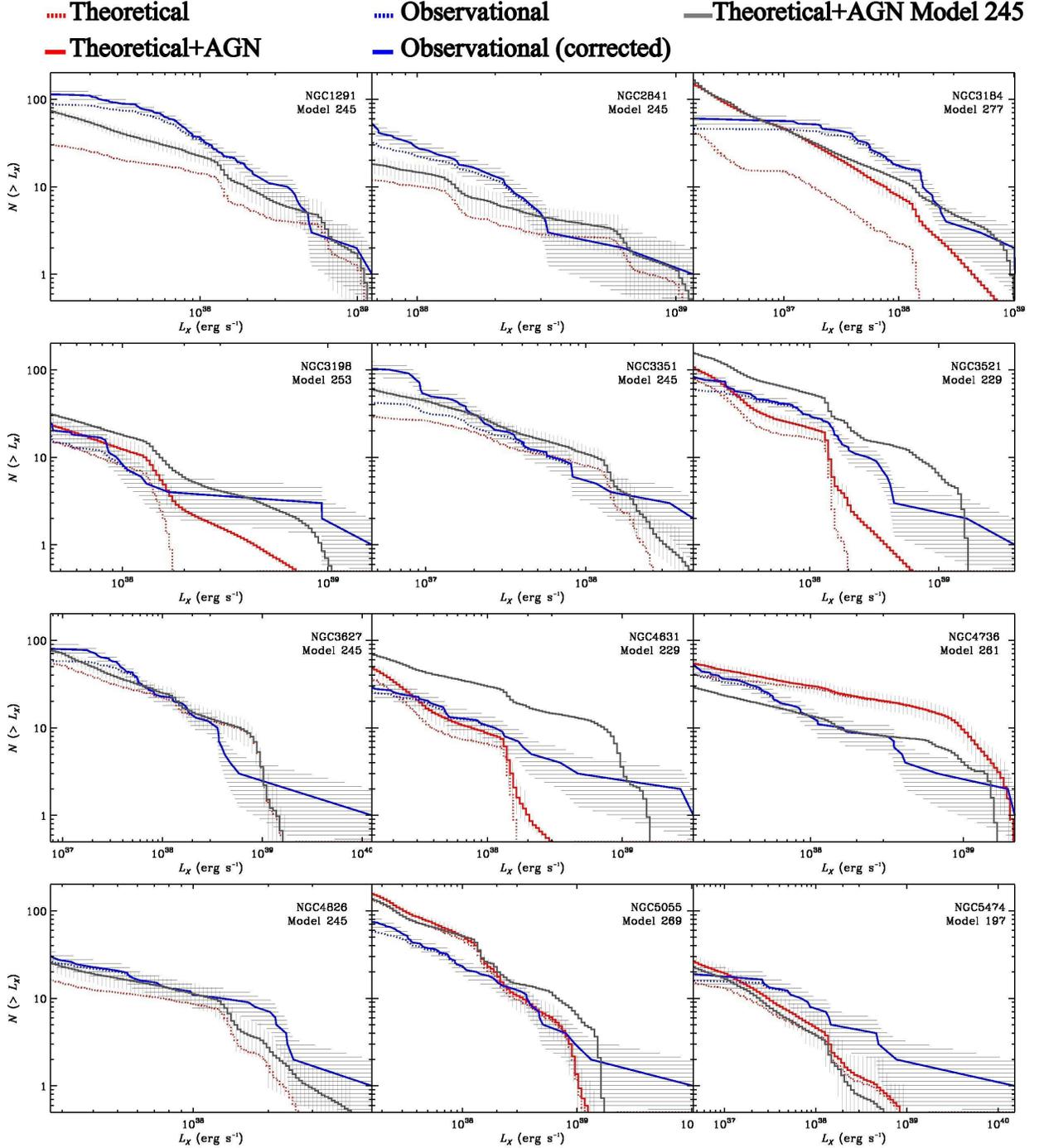}
\caption{
Observational and theoretical cumulative XLFs for SINGS galaxies.
Blue: observational (dotted, original; solid, corrected for
incompleteness). Red: theoretical XLF that has
highest {\it pair} likelihood 
(\tr{tab-bestgalmod})
for a given galaxy (dotted, original; solid,
with added expected unrelated contribution from background AGN). 
Dark grey: theoretical
XLF (with added expected AGN contribution) 
for the galaxy shown, but using \stk\ model 245, which
has the highest {\it global} likelihood.
For galaxies 
NGC 1291, 2841, 3351, 3627, and 4826
the solid red and solid dark grey
curves coincide.
The horizontal (vertical) hashed regions indicate $\pm 1\sigma$ estimates for
the oXLF and tXLF, respectively, due to Poisson statistics
and bolometric uncertainties.
\label{fig_otxlf}}
\end{figure*}
 

\begin{deluxetable*}{cccc}
\scriptsize
\tablecolumns{4}
\tablewidth{500 pt}
\setlength{\tabcolsep}{.1 pt} 
\tablecaption{Model Parameters. \label{tab-param}}
\tablehead{ 
\colhead{Parameter\tablenotemark{a}} &\colhead{Notation} &\colhead{Value} &\colhead{Reference}
}
\startdata
Initial Orbital Period distribution   &     $F(P)$ & \bf{flat in log}~$\bm{a}$ \tablenotemark{b}& \citet{abt1983} \\
Initial Eccentricity Distribution     &     $F(e)$ & \bf{Thermal} $\bm{ F(e) \sim e}$     & \citet{heggie1975} \\
Binary Fraction                       &$f_{\rm bin}$& \bf{50\%}                            & \\
Magnetic Braking                      &            &                                      & \citet{ivanova2003} \\
Metallicity                           & $Z$        & 0.0001, 0.0002, 0.005, 0.001,        & \\
                                      &            & 0.002, 0.005, 0.01, {\bf 0.02}, 0.03 & \\
IMF high-end slope                    &            & {\bf -2.35} or {\bf -2.7}                   & \citet{kroupa2001}; \citet{kroupa2003} \\
Initial Mass Ratio Distribution       & $F(q)$     &   {\bf Flat}, twin, or {\bf 50\% flat -- 50\% twin}&    \citet{kobulnicky2007}; \citet{pinsonneault2006}\\
CE Efficiency$\times$ central concentration & \acol\      & {\bf 0.1, 0.2, 0.3, 0.5}     & \citet{podsiadlowski2003} \\
Stellar wind strength                 & \ewind\    & {\bf 0.25, 1.0, 2.0}         & \citet{belczynski2010} \\
CE during HG                          &  CE-HG     & {\bf Yes} or {\bf No}              & \citet{belczynski2007} \\
SN kick for ECS/AIC\tablenotemark{c}  NS &         & {\bf 20\% of normal NS kicks} & \citet{linden2009}    \\
SN kick for direct collapse BH\tablenotemark{d}       & \kdcbh\    & {\bf Yes (0.1)} or {\bf No (0)}               & \citet{fragos2010}    \\
\enddata
\tablenotetext{a}{The full range of values shown in column 3 is shown only for reference. Although our model grid
contains results for these values, {\it in this paper} we only use models with the values shown in bold in
column 3. As explained in the text, this corresponds to using (1) only models $1-96$ (flat $q$ distribution) and $193-288$
(50\%\ -- 50\%\ $q$ distribution), and (2) only solar metallicities in all cases.}
\tablenotetext{b}{$a$ is the semi-major axis of the binary orbit.}
\tablenotetext{c}{Electron Capture Supernova / Accretion Induced Collapse}
\tablenotetext{d}{The Hobbs et al. (2005) kick distribution for NS is multiplied by this parameter to introduce small
kicks for BHs formed through a SN explosion with negligible ejected mass.}
\end{deluxetable*}

\begin{figure*}
\epsscale{1.0}
\plotone{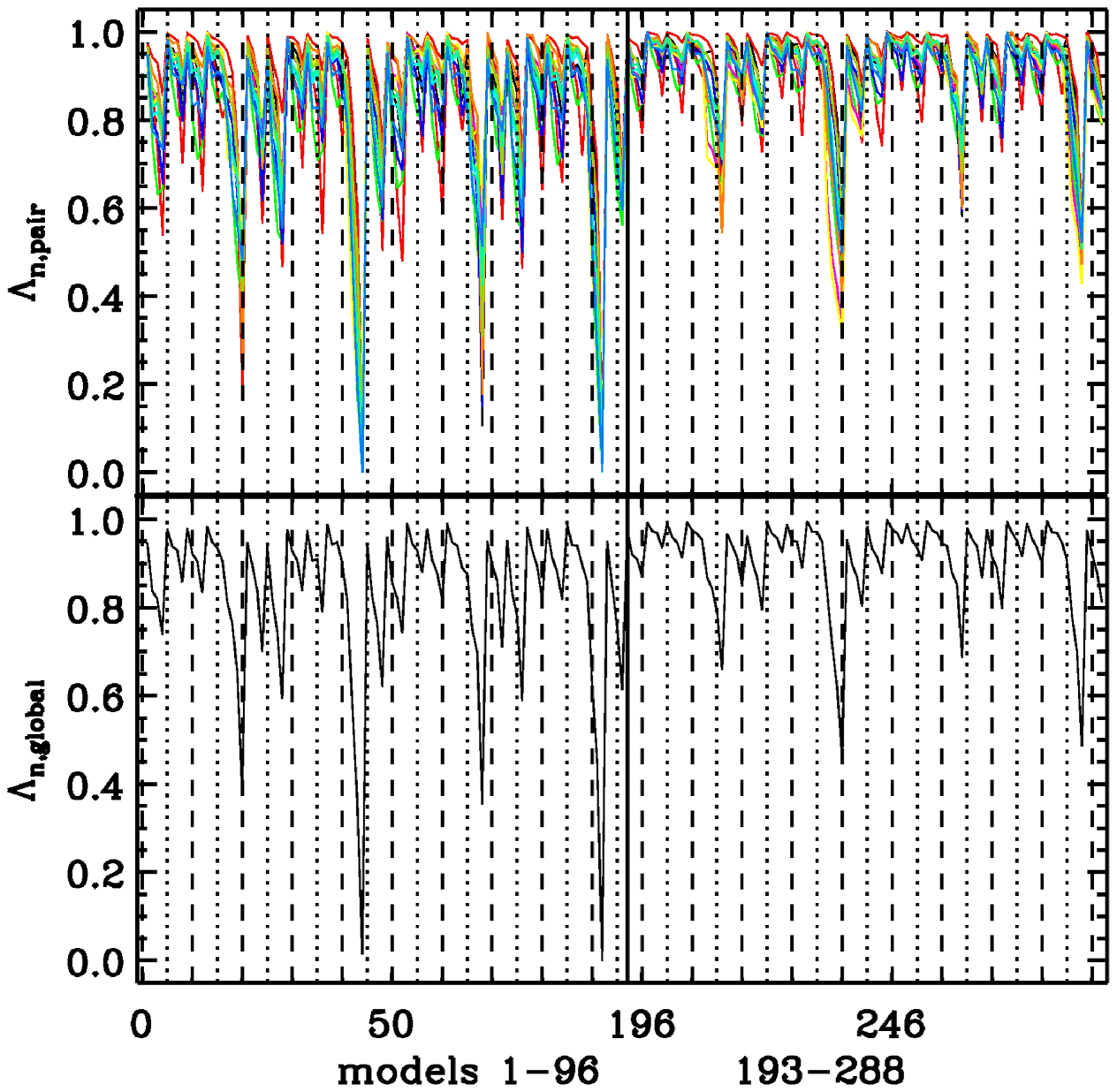}
\caption{
Normalized pair (top) and global (bottom)
likelihood (\scr{sec_lnorm}) 
vs. model number for all models used in this paper.
As explained in the text, the global curve in the bottom
panel is
the combination of the individual curves in the top panel.
\label{fig_lik_pair_glob_mods}}
\end{figure*}

\begin{figure*}
\epsscale{0.5}
\plotone{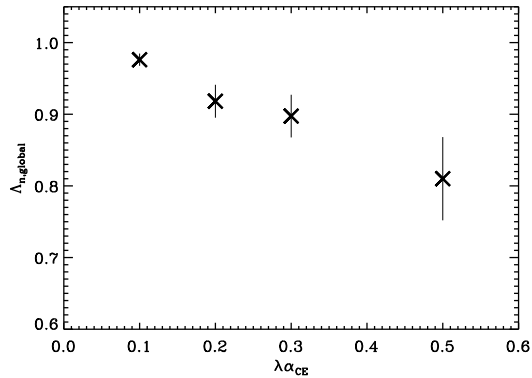}
\caption{Normalized global likelihood (\scr{sec_lnorm})
vs. \stk\ parameter \acol\
(donor central concentration $\times$
common envelope efficiency) varied
in our model grid. Crosses indicate resistant mean values
for normalized global likelihood values of models
using a given \acol\ value. Error bars indicate
$3\sigma$ estimates.
\label{fig_likparam}}
\end{figure*}

 

\begin{figure*}
\epsscale{1.1}
\plotone{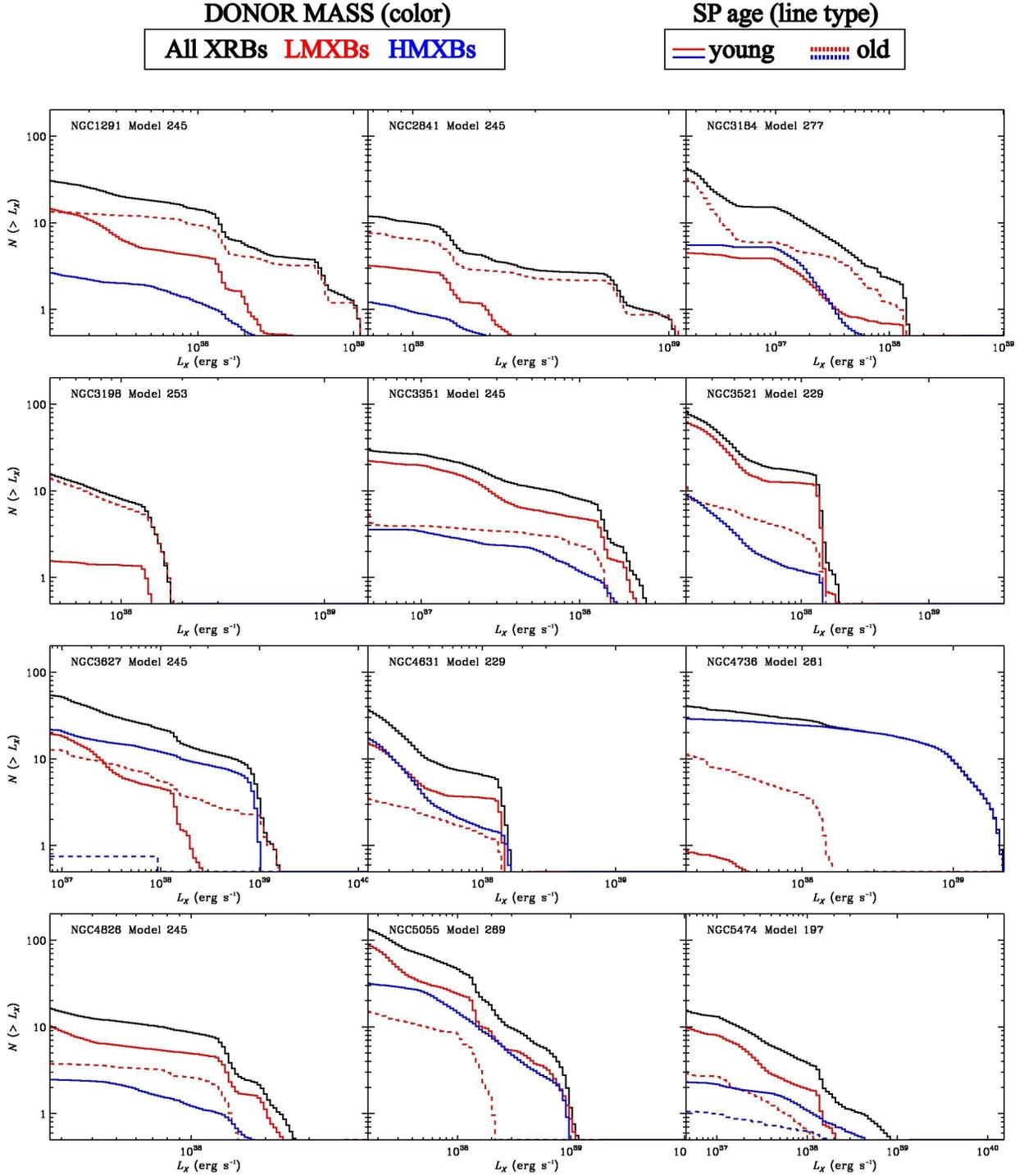}
\caption{Cumulative theoretical XLFs for XSINGS galaxies, for subpopulations
based on donor mass and age of stellar population.
Continuous black curves show the total tXLF (black). 
The \lq\lq old\rq\rq\ LMXB population is shown by the red dotted
lines, and the \lq\lq old\rq\rq\ HMXB population by the blue
dotted line.
The \lq\lq young\rq\rq\ LMXB population is shown by the red solid line,
and the \lq\lq\ young\rq\rq\ HMXB population by the blue solid line.
The \lq\lq young\rq\rq\ and \lq\lq old\rq\rq\ subpopulations
are defined in \citet{noll2009} and are shown separately
by solid and dotted lines, respectively (see also \tr{tab-sfh}).
Note that the \lx\ axes have different ranges, matching the
observed XLF range.
\label{fig_mass_oy}}
\end{figure*}

\begin{figure*}
\epsscale{1.1}
\plotone{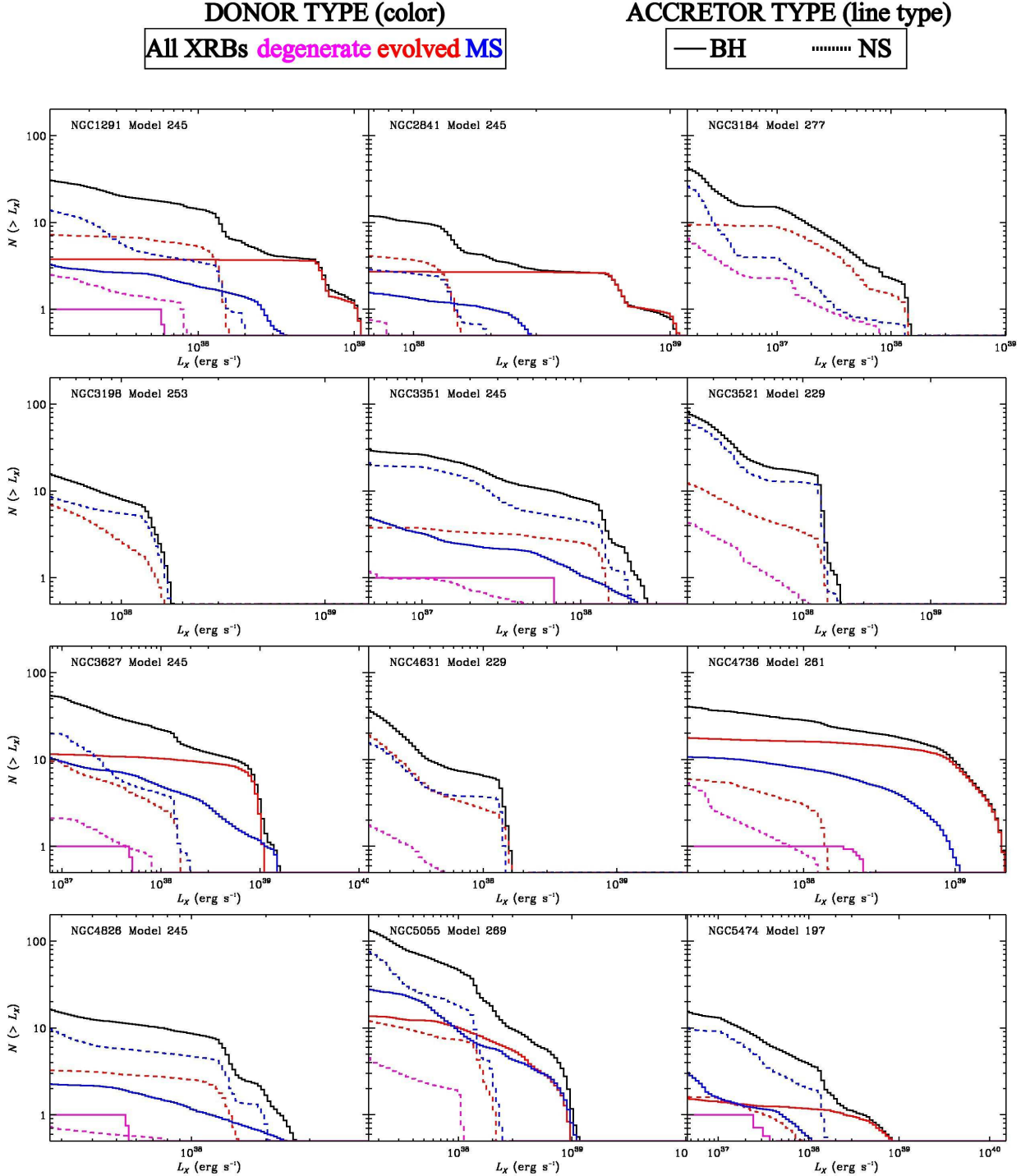}
\caption{Cumulative theoretical XLFs for XSINGS galaxies, 
for subpopulations based on donor and accretor type.
The continuous black curve is the total tXLF.
For colored curves,
color indicates donor stellar type
and line type indicates accretor type, as indicated at the top
of the figure.
\label{fig_da}}
\end{figure*}

\begin{figure*}
\epsscale{1.0}
\plottwo{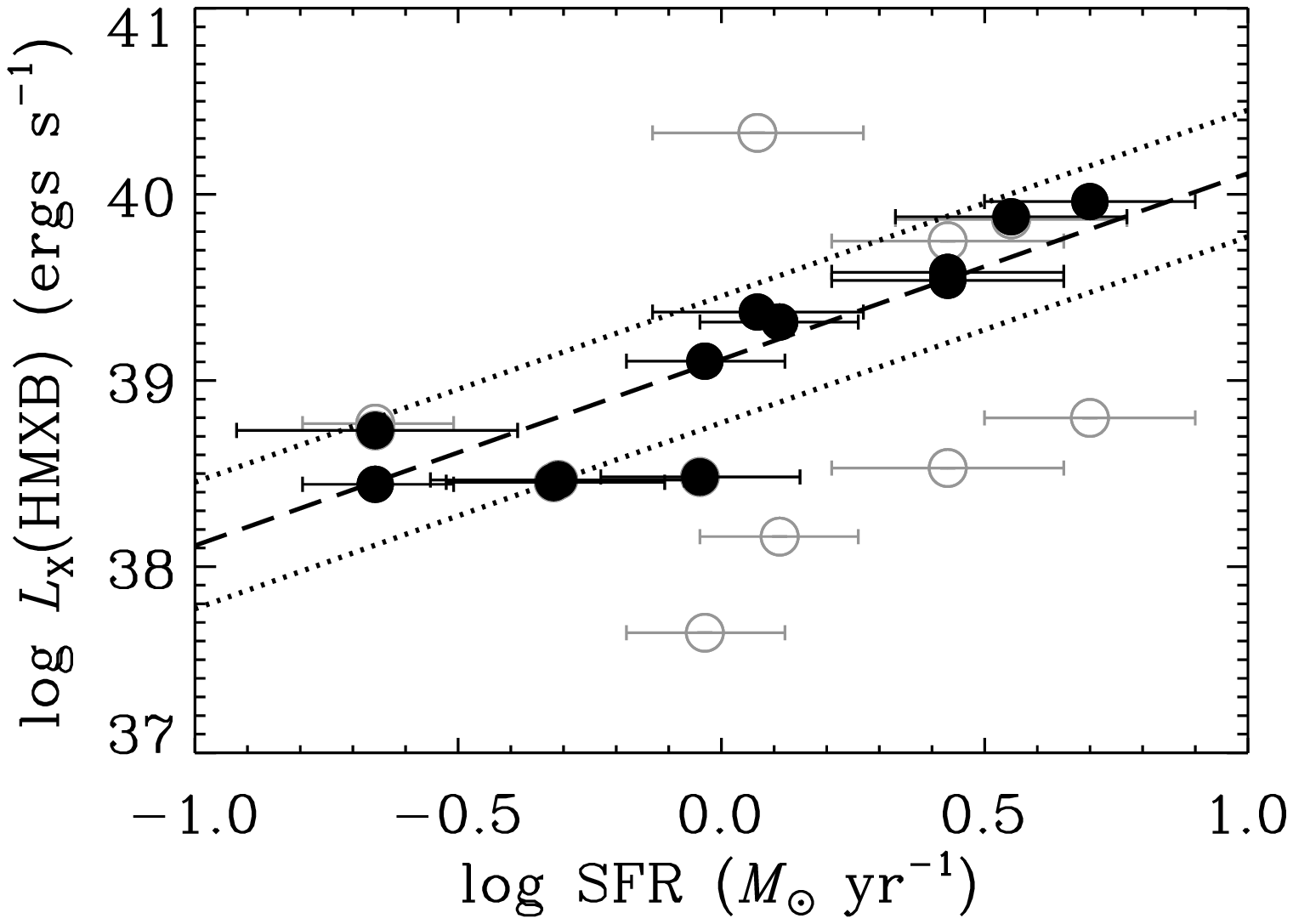}{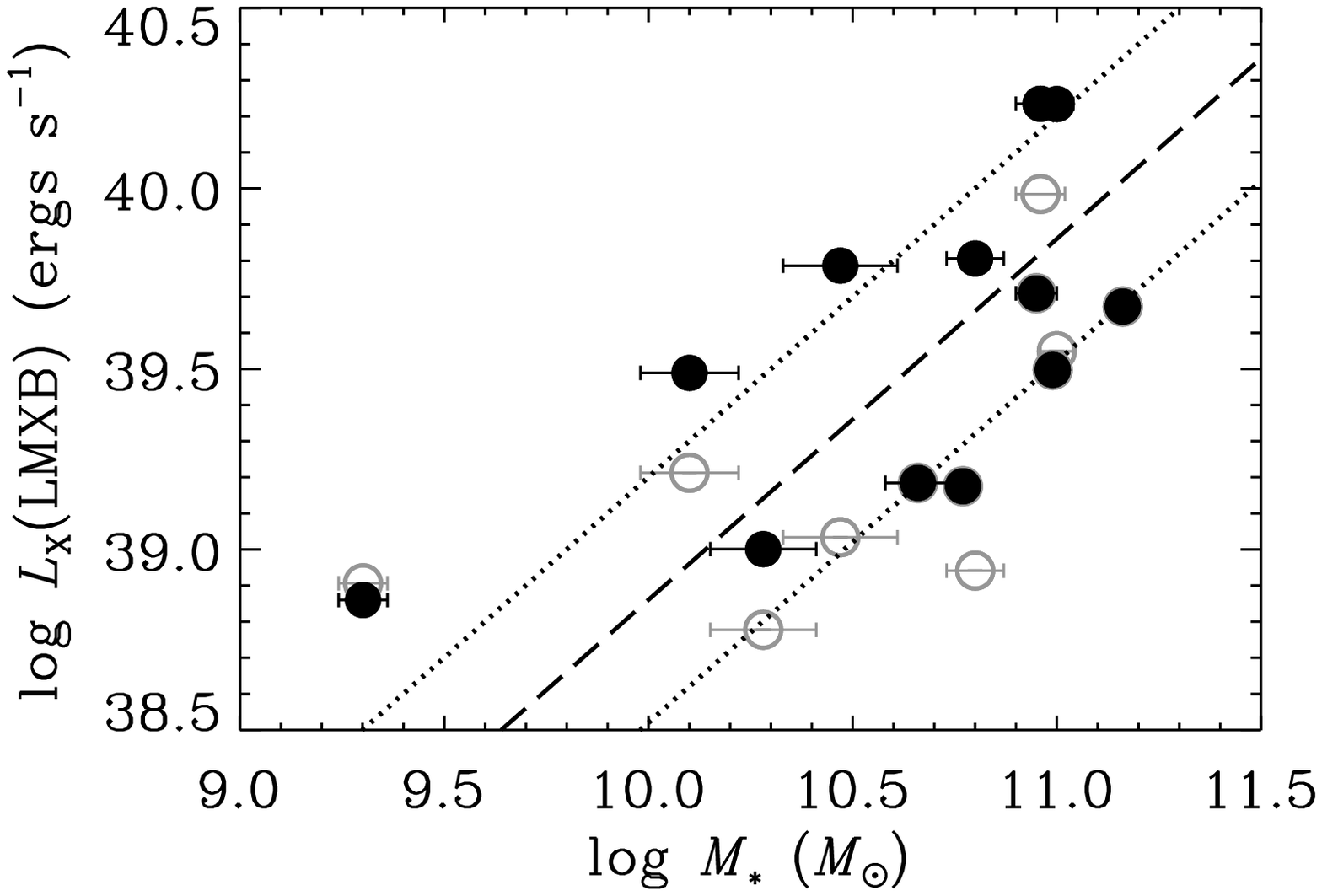}
\caption{Integrated \lxtt\ from LMXBs and HMXBs from 
\stk\ modeling vs. SFR and \mstar\ \citep{noll2009} for
the 12 SINGS galaxies in this paper. 
Luminosities have been calculated using both the
best pair-likelihood model for each galaxy (open grey
circles) and model 245, the best global model in this paper,
which is also
the top-ranked model in F13 (black filled circles).
The dashed line
shows the expected contribution from HMXBs (left panel) and
LMXBs (right panel) based on the best-fit relation of L10 (\er{equ:L10}).
The dotted lines show $\pm1\sigma$ uncertainties.
\label{fig_L10}}
\end{figure*}


 \begin{deluxetable*}{cccc cccc}
\tablecolumns{9}
\tablewidth{0pc} 
\tablecaption{SINGS galaxy subsample used in this paper.\label{tab-sample}}
\tablehead{ 
\colhead{Galaxy}
&\colhead{$D$}
&\colhead{Hubble Type}
&\colhead{T-Type}
&\colhead{log SFR}
&\colhead{log \mstar}
&\colhead{SSFR}
&\colhead{$N_{\rm src}$}
\\
\colhead{}
&\colhead{(Mpc)}
&\colhead{}
&\colhead{}
&\colhead{(\msuny)}
&\colhead{(\msun)}
&\colhead{($10^{-11}$ yr$^{-1}$)}
&\colhead{}
\\
\colhead{(1)}
&\colhead{(2)}
&\colhead{(3)}
&\colhead{(4)}
&\colhead{(5)}
&\colhead{(6)}
&\colhead{(7)}
&\colhead{(8)}
}
\startdata
        NGC1291              &           10.8              &            SBa              &              1              & $-0.66\pm0.27$              & $11.16\pm0.03$              &       0.151356              &             88              \\
        NGC2841              &           14.1              &            SAb              &              3              & $-0.31\pm0.24$              & $10.99\pm0.03$              &       0.501187              &             32              \\
        NGC3184              &           11.1              &          SABcd              &              6              &  $0.11\pm0.15$              & $10.28\pm0.13$              &        6.76083              &             47              \\
        NGC3198              &          13.68              &            SBc              &              5              & $-0.03\pm0.15$              & $10.10\pm0.12$              &         7.4131              &             19              \\
        NGC3351              &           9.33              &            SBb              &              3              & $-0.04\pm0.19$              & $10.66\pm0.08$              &        1.99526              &             45              \\
        NGC3521              &           10.1              &          SABbc              &              4              &  $0.43\pm0.22$              & $11.00\pm0.04$              &        2.69153              &             60              \\
        NGC3627              &           9.38              &           SABb              &              3              &  $0.55\pm0.22$              & $10.95\pm0.05$              &        3.98107              &             59              \\
        NGC4631              &           7.62              &            SBd              &              7              &  $0.70\pm0.20$              & $10.47\pm0.14$              &        16.9824              &             29              \\
        NGC4736              &            5.2              &           SAab              &              2              &  $0.07\pm0.20$              & $10.80\pm0.07$              &        1.86209              &             42              \\
        NGC4826              &           7.48              &           SAab              &              2              & $-0.32\pm0.21$              & $10.77\pm0.02$              &       0.812831              &             28              \\
        NGC5055              &            7.8              &           SAbc              &              4              &  $0.43\pm0.22$              & $10.96\pm0.06$              &        2.95121              &             59              \\
        NGC5474              &            6.8              &           SAcd              &              6              & $-0.65\pm0.14$              &  $9.30\pm0.06$              &        11.2202              &             16              \\

\enddata
\tablecomments{Columns are: (1) galaxy name; (2) distance \citep{moustakas2010}; (3) Hubble Type \citep{devaucouleurs1991}; (4) T-type; (5) star formation rate from \citet{noll2009}; (6) logarithmic stellar mass from \citet{noll2009}; (7) specific star formation rate (SFR/\mstar) from (5) and (6); 
(8) number of \x\ point sources in XSINGS catalog.}
\end{deluxetable*}

 \begin{deluxetable*}{cccc ccc}
\tablecolumns{7}
\tablewidth{0pc} 
\tablecaption{Star formation histories from SED fitting. \label{tab-sfh}}
\tablehead{ 
\colhead{Galaxy}
&\colhead{log \mgal}
&\colhead{$\log \tau_{\rm oSP}$}
&\colhead{$t_{\rm oSP}$}
&\colhead{$\log \tau_{\rm ySP}$}
&\colhead{$t_{\rm ySP}$}
&\colhead{$\log f_{\rm burst}$}
\\
\colhead{}
&\colhead{(\msun)}
&\colhead{(Gyr)}
&\colhead{(Gyr)}
&\colhead{(Gyr)}
&\colhead{(Gyr)}
&\colhead{}
\\
\colhead{(1)}
&\colhead{(2)}
&\colhead{(3)}
&\colhead{(4)}
&\colhead{(5)}
&\colhead{(6)}
&\colhead{(7)}
}
\startdata
        NGC1291 & 11.3  & -0.60 & 10.0 & -1.30 & 0.20 & -2.52 \\
        NGC2841 & 11.2  & -0.60 & 10.0 & -1.30 & 0.20 & -2.52 \\
        NGC3184 & 10.6  & -0.60 & 10.0 &  0.00 & 0.20 & -2.00 \\
        NGC3198 & 10.1  &  1.00 & 10.0 & -1.30 & 0.20 & -1.52 \\
        NGC3351 & 10.9  & -0.60 & 10.0 & -1.30 & 0.20 & -2.00 \\
        NGC3521 & 11.2  & -0.60 & 10.0 & -1.30 & 0.20 & -1.52 \\
        NGC3627 & 11.1  &  0.00 & 10.0 &  0.00 & 0.20 & -2.00 \\
        NGC4631 & 10.8  & -0.60 & 10.0 & -0.60 & 0.20 & -1.52 \\
        NGC4736 & 11.0  & -0.60 & 10.0 &  0.00 & 0.05 & -3.00 \\
        NGC4826 & 10.9  & -0.60 & 10.0 & -1.30 & 0.20 & -2.00 \\
        NGC5055 & 11.2  & -0.60 & 10.0 & -1.30 & 0.20 & -1.52 \\
        NGC5474 &  9.5  &  0.48 & 10.0 & -1.30 & 0.20 & -1.00 \\

\enddata
\tablecomments{Best fit values from SED fitting (run B in \citet{noll2009} and 
Noll, private communication). Columns are: (1) galaxy name; (2) total stellar mass plus gas mass from stellar mass loss; (3) e-folding timescale for old stellar population; (4) age of old stellar population; (5) e-folding timescale for young stellar population; (6) age of young stellar population; (7) mass fraction of young stellar population.
}
\end{deluxetable*}


\begin{deluxetable*}{cccc cccc cc}
\tablecolumns{9}
\tablewidth{0pc} 
\tablecaption{Parameters for highest pair-likelihood model-galaxy pairs \label{tab-bestgalmod}}
\tablehead{ 
\colhead{Galaxy}
&\colhead{Model}
&\colhead{\acol}
&\colhead{IMF exponent}
&\colhead{\ewind}
&\colhead{CE-HG}
&\colhead{$q$ distribution}
&\colhead{\kdcbh}
&\colhead{Rank}
&\colhead{likelihood ratio}
\\
\colhead{(1)}
&\colhead{(2)}
&\colhead{(3)}
&\colhead{(4)}
&\colhead{(5)}
&\colhead{(6)}
&\colhead{(7)}
&\colhead{(8)}
&\colhead{(9)}
&\colhead{(10)}
}
\startdata
NGC1291  &  245  &  0.1  &  -2.7  &  1.0  &  No  &  50-50  &  0.1  & 	10	 &  -381 \\
NGC2841  &  245  &  0.1  &  -2.7  &  1.0  &  No  &  50-50  &  0.1  & 	1	 &     0 \\
NGC3184  &  277  &  0.1  &  -2.7  &  2.0  &  Yes  &  50-50  &  0.1  & 	12	 &  5636 \\
NGC3198  &  253  &  0.1  &  -2.7  &  2.0  &  No  &  50-50  &  0.1  & 	2	 &    -7 \\
NGC3351  &  245  &  0.1  &  -2.7  &  1.0  &  No  &  50-50  &  0.1  & 	11	 &  -425 \\
NGC3521  &  229  &  0.1  &  -2.7  &  2.0  &  Yes  &  50-50  &  0  & 	9	 &  -180 \\
NGC3627  &  245  &  0.1  &  -2.7  &  1.0  &  No  &  50-50  &  0.1  & 	8	 &  -177 \\
NGC4631  &  229  &  0.1  &  -2.7  &  2.0  &  Yes  &  50-50  &  0  & 	5	 &   -44 \\
NGC4736  &  261  &  0.1  &  -2.7  &  0.25  &  No  &  50-50  &  0.1  & 	4	 &   -23 \\
NGC4826  &  245  &  0.1  &  -2.7  &  1.0  &  No  &  50-50  &  0.1  & 	3	 &    -8 \\
NGC5055  &  269  &  0.1  &  -2.7  &  1.0  &  Yes  &  50-50  &  0.1  & 	7	 &   -84 \\
NGC5474  &  197  &  0.1  &  -2.7  &  1.0  &  No  &  50-50  &  0  & 	6	 &   -45 \\

\enddata
\tablecomments{Columns are: (1) galaxy ID; (2) model number; (3) CE efficiency
$\times$ central concentration; (4) exponent of high-mass power law component
of IMF: Kroupa (2001, $-2.35$) or Kroupa \& Weidner (2003, $-2.7$);
(5) stellar wind strength
parameter; (6) Yes: All possible outcomes of a CE event with
a Hertzsprung gap donor allowed; No: A CE with such a donor star will 
always result to a merger; (7) binary mass ratio distribution: 50-50
means half of the binaries originate in a twin binary distribution
and half in a flat mass ratio distribution; (8) parameter with
which the Hobbs et al. (2005) kick distribution is multiplied for
BHs formed through a SN explosion with negligible ejected mass; 
(9) rank of best-fit galaxy-model pairs based on pair likelihood value;
(10) natural logarithm of ratio of each pair likelihood to maximum pair
likelihood in this table:
$\ln ( {\cal L}_{{\rm pair,} km}/ {\cal L}_{{\rm pair,} km, {\rm max}}) $.}
\end{deluxetable*}

\begin{deluxetable*}{cccc cccc ccc}
\tablecolumns{11}
\tablewidth{0pc} 
\tablecaption{Parameters and global likelihood values for 15 best models ranked by likelihood. \label{tab-bestbymod15}}
\tablehead{ 
\colhead{Model}
&\colhead{\acol}
&\colhead{IMF exponent}
&\colhead{\ewind}
&\colhead{CE-HG}
&\colhead{$q$ distribution}
&\colhead{\kdcbh}
&\colhead{Rank}
&\colhead{Rank F13}
&\colhead{Rank T13}
&\colhead{likelihood ratio}
\\
\colhead{(1)}
&\colhead{(2)}
&\colhead{(3)}
&\colhead{(4)}
&\colhead{(5)}
&\colhead{(6)}
&\colhead{(7)}
&\colhead{(8)}
&\colhead{(9)}
&\colhead{(10)}
&\colhead{(11)}
}
\startdata
245 & 0.1 & -2.7 & 1.0 & No & 50-50 & 0.1 & 1 & 1 & 4 &  0 \\
253 & 0.1 & -2.7 & 2.0 & No & 50-50 & 0.1 & 2 & 14 & 5 &  -81 \\
277 & 0.1 & -2.7 & 2.0 & Yes & 50-50 & 0.1 & 3 & 8 & 3 &  -97 \\
229 & 0.1 & -2.7 & 2.0 & Yes & 50-50 & 0 & 4 & 2 & 2 &  -128 \\
269 & 0.1 & -2.7 & 1.0 & Yes & 50-50 & 0.1 & 5 & 3 & 12 &  -137 \\
205 & 0.1 & -2.7 & 2.0 & No & 50-50 & 0 & 6 & 4 & 1 &  -198 \\
197 & 0.1 & -2.7 & 1.0 & No & 50-50 & 0 & 7 & 13 & 50 &  -207 \\
61 & 0.1 & -2.7 & 2.0 & No & Flat & 0.1 & 8 & 16 & 7 &  -258 \\
201 & 0.1 & -2.35 & 2.0 & No & 50-50 & 0 & 9 & 10 & 19 &  -260 \\
221 & 0.1 & -2.7 & 1.0 & Yes & 50-50 & 0 & 10 & 15 & 60 &  -262 \\
53 & 0.1 & -2.7 & 1.0 & No & Flat & 0.1 & 11 & 20 & 22 &  -283 \\
273 & 0.1 & -2.35 & 2.0 & Yes & 50-50 & 0.1 & 12 & 6 & 6 &  -285 \\
249 & 0.1 & -2.35 & 2.0 & No & 50-50 & 0.1 & 13 & 5 & 10 &  -296 \\
85 & 0.1 & -2.7 & 2.0 & Yes & Flat & 0.1 & 14 & 12 & 9 &  -305 \\
37 & 0.1 & -2.7 & 2.0 & Yes & Flat & 0 & 15 & 7 & 8 &  -341 \\

\enddata
\tablecomments{Columns are: (1) model number; (2) CE efficiency
$\times$ central concentration; (3) exponent of high-mass power law component
of IMF: Kroupa (2001, $-2.35$) or Kroupa \& Weidner (2003, $-2.7$);
(4) stellar wind strength
parameter; (5) Yes: All possible outcomes of a CE event with
a Hertzsprung gap donor allowed; No: A CE with such a donor star will 
always result to a merger; (6) binary mass ratio distribution: 50-50
means half of the binaries originate in a twin binary distribution
and half in a flat mass ratio distribution; (7) parameter with
which the Hobbs et al. (2005) kick distribution is multiplied for
BHs formed through a SN explosion with negligible ejected mass; 
(8) rank of model based on likelihood value in this paper; 
(9) rank of model based on likelihood value in F13;
(10) rank of model based on likelihood value in T13;
(11) 
natural logarithm of ratio of each global likelihood to
maximum global likelihood in this table:
$\ln(\cal L_{\rm global}/\cal L_{\rm global, max})$.
(The full table is available on-line).}
\end{deluxetable*}

\clearpage
\LongTables

\begin{deluxetable*}{cccc cccc ccc}
\tablecolumns{11}
\tablewidth{0pc} 
\tablecaption{Parameters and global likelihood values for 192 models ranked by likelihood ({\bf on-line only}). \label{tab-bestbymod15}}
\tablehead{ 
\colhead{Model}
&\colhead{\acol}
&\colhead{IMF exponent}
&\colhead{\ewind}
&\colhead{CE-HG}
&\colhead{$q$ distribution}
&\colhead{\kdcbh}
&\colhead{Rank}
&\colhead{Rank F13}
&\colhead{Rank T13}
&\colhead{likelihood ratio}
\\
\colhead{(1)}
&\colhead{(2)}
&\colhead{(3)}
&\colhead{(4)}
&\colhead{(5)}
&\colhead{(6)}
&\colhead{(7)}
&\colhead{(8)}
&\colhead{(9)}
&\colhead{(10)}
&\colhead{(11)}
\\
\colhead{}
&\colhead{}
&\colhead{}
&\colhead{}
&\colhead{}
&\colhead{}
&\colhead{}
&\colhead{}
&\colhead{}
&\colhead{}
&\colhead{}
}
\startdata
245 & 0.1 & -2.7 & 1.0 & No & 50-50 & 0.1 & 1 & 1 & 4 &  0 \\
253 & 0.1 & -2.7 & 2.0 & No & 50-50 & 0.1 & 2 & 14 & 5 &  -81 \\
277 & 0.1 & -2.7 & 2.0 & Yes & 50-50 & 0.1 & 3 & 8 & 3 &  -97 \\
229 & 0.1 & -2.7 & 2.0 & Yes & 50-50 & 0 & 4 & 2 & 2 &  -128 \\
269 & 0.1 & -2.7 & 1.0 & Yes & 50-50 & 0.1 & 5 & 3 & 12 &  -137 \\
205 & 0.1 & -2.7 & 2.0 & No & 50-50 & 0 & 6 & 4 & 1 &  -198 \\
197 & 0.1 & -2.7 & 1.0 & No & 50-50 & 0 & 7 & 13 & 50 &  -207 \\
61 & 0.1 & -2.7 & 2.0 & No & Flat & 0.1 & 8 & 16 & 7 &  -258 \\
201 & 0.1 & -2.35 & 2.0 & No & 50-50 & 0 & 9 & 10 & 19 &  -260 \\
221 & 0.1 & -2.7 & 1.0 & Yes & 50-50 & 0 & 10 & 15 & 60 &  -262 \\
53 & 0.1 & -2.7 & 1.0 & No & Flat & 0.1 & 11 & 20 & 22 &  -283 \\
273 & 0.1 & -2.35 & 2.0 & Yes & 50-50 & 0.1 & 12 & 6 & 6 &  -285 \\
249 & 0.1 & -2.35 & 2.0 & No & 50-50 & 0.1 & 13 & 5 & 10 &  -296 \\
85 & 0.1 & -2.7 & 2.0 & Yes & Flat & 0.1 & 14 & 12 & 9 &  -305 \\
37 & 0.1 & -2.7 & 2.0 & Yes & Flat & 0 & 15 & 7 & 8 &  -341 \\
225 & 0.1 & -2.35 & 2.0 & Yes & 50-50 & 0 & 16 & 11 & 33 &  -358 \\
241 & 0.1 & -2.35 & 1.0 & No & 50-50 & 0.1 & 17 & 31 & 59 &  -452 \\
13 & 0.1 & -2.7 & 2.0 & No & Flat & 0 & 18 & 9 & 11 &  -473 \\
77 & 0.1 & -2.7 & 1.0 & Yes & Flat & 0.1 & 19 & 21 & 27 &  -542 \\
261 & 0.1 & -2.7 & 0.25 & No & 50-50 & 0.1 & 20 & 43 & 81 &  -577 \\
9 & 0.1 & -2.35 & 2.0 & No & Flat & 0 & 21 & 29 & 29 &  -608 \\
57 & 0.1 & -2.35 & 2.0 & No & Flat & 0.1 & 22 & 28 & 30 &  -652 \\
246 & 0.2 & -2.7 & 1.0 & No & 50-50 & 0.1 & 23 & 25 & 23 &  -675 \\
81 & 0.1 & -2.35 & 2.0 & Yes & Flat & 0.1 & 24 & 26 & 20 &  -680 \\
5 & 0.1 & -2.7 & 1.0 & No & Flat & 0 & 25 & 23 & 42 &  -682 \\
29 & 0.1 & -2.7 & 1.0 & Yes & Flat & 0 & 26 & 24 & 48 &  -706 \\
265 & 0.1 & -2.35 & 1.0 & Yes & 50-50 & 0.1 & 27 & 41 & 79 &  -726 \\
254 & 0.2 & -2.7 & 2.0 & No & 50-50 & 0.1 & 28 & 22 & 13 &  -756 \\
285 & 0.1 & -2.7 & 0.25 & Yes & 50-50 & 0.1 & 29 & 46 & 87 &  -757 \\
206 & 0.2 & -2.7 & 2.0 & No & 50-50 & 0 & 30 & 18 & 14 &  -757 \\
193 & 0.1 & -2.35 & 1.0 & No & 50-50 & 0 & 31 & 56 & 109 &  -777 \\
33 & 0.1 & -2.35 & 2.0 & Yes & Flat & 0 & 32 & 30 & 32 &  -782 \\
198 & 0.2 & -2.7 & 1.0 & No & 50-50 & 0 & 33 & 39 & 65 &  -802 \\
231 & 0.3 & -2.7 & 2.0 & Yes & 50-50 & 0 & 34 & 37 & 31 &  -831 \\
230 & 0.2 & -2.7 & 2.0 & Yes & 50-50 & 0 & 35 & 17 & 16 &  -840 \\
278 & 0.2 & -2.7 & 2.0 & Yes & 50-50 & 0.1 & 36 & 19 & 15 &  -872 \\
270 & 0.2 & -2.7 & 1.0 & Yes & 50-50 & 0.1 & 37 & 34 & 39 &  -929 \\
255 & 0.3 & -2.7 & 2.0 & No & 50-50 & 0.1 & 38 & 40 & 21 &  -933 \\
199 & 0.3 & -2.7 & 1.0 & No & 50-50 & 0 & 39 & 55 & 70 &  -935 \\
279 & 0.3 & -2.7 & 2.0 & Yes & 50-50 & 0.1 & 40 & 33 & 24 &  -944 \\
247 & 0.3 & -2.7 & 1.0 & No & 50-50 & 0.1 & 41 & 45 & 40 &  -946 \\
222 & 0.2 & -2.7 & 1.0 & Yes & 50-50 & 0 & 42 & 42 & 71 &  -990 \\
207 & 0.3 & -2.7 & 2.0 & No & 50-50 & 0 & 43 & 32 & 25 &  -997 \\
213 & 0.1 & -2.7 & 0.25 & No & 50-50 & 0 & 44 & 84 & 111 &  -1009 \\
217 & 0.1 & -2.35 & 1.0 & Yes & 50-50 & 0 & 45 & 60 & 115 &  -1111 \\
49 & 0.1 & -2.35 & 1.0 & No & Flat & 0.1 & 46 & 58 & 74 &  -1215 \\
202 & 0.2 & -2.35 & 2.0 & No & 50-50 & 0 & 47 & 48 & 45 &  -1227 \\
223 & 0.3 & -2.7 & 1.0 & Yes & 50-50 & 0 & 48 & 57 & 85 &  -1297 \\
274 & 0.2 & -2.35 & 2.0 & Yes & 50-50 & 0.1 & 49 & 47 & 41 &  -1339 \\
262 & 0.2 & -2.7 & 0.25 & No & 50-50 & 0.1 & 50 & 87 & 98 &  -1357 \\
237 & 0.1 & -2.7 & 0.25 & Yes & 50-50 & 0 & 51 & 91 & 113 &  -1398 \\
208 & 0.5 & -2.7 & 2.0 & No & 50-50 & 0 & 52 & 27 & 18 &  -1400 \\
73 & 0.1 & -2.35 & 1.0 & Yes & Flat & 0.1 & 53 & 71 & 88 &  -1430 \\
226 & 0.2 & -2.35 & 2.0 & Yes & 50-50 & 0 & 54 & 54 & 62 &  -1441 \\
250 & 0.2 & -2.35 & 2.0 & No & 50-50 & 0.1 & 55 & 44 & 36 &  -1448 \\
271 & 0.3 & -2.7 & 1.0 & Yes & 50-50 & 0.1 & 56 & 53 & 56 &  -1464 \\
93 & 0.1 & -2.7 & 0.25 & Yes & Flat & 0.1 & 57 & 86 & 107 &  -1464 \\
69 & 0.1 & -2.7 & 0.25 & No & Flat & 0.1 & 58 & 79 & 103 &  -1484 \\
227 & 0.3 & -2.35 & 2.0 & Yes & 50-50 & 0 & 59 & 80 & 73 &  -1541 \\
232 & 0.5 & -2.7 & 2.0 & Yes & 50-50 & 0 & 60 & 35 & 28 &  -1551 \\
21 & 0.1 & -2.7 & 0.25 & No & Flat & 0 & 61 & 101 & 119 &  -1563 \\
14 & 0.2 & -2.7 & 2.0 & No & Flat & 0 & 62 & 51 & 37 &  -1575 \\
39 & 0.3 & -2.7 & 2.0 & Yes & Flat & 0 & 63 & 76 & 57 &  -1581 \\
62 & 0.2 & -2.7 & 2.0 & No & Flat & 0.1 & 64 & 52 & 34 &  -1586 \\
203 & 0.3 & -2.35 & 2.0 & No & 50-50 & 0 & 65 & 66 & 64 &  -1598 \\
1 & 0.1 & -2.35 & 1.0 & No & Flat & 0 & 66 & 81 & 105 &  -1610 \\
280 & 0.5 & -2.7 & 2.0 & Yes & 50-50 & 0.1 & 67 & 36 & 26 &  -1619 \\
248 & 0.5 & -2.7 & 1.0 & No & 50-50 & 0.1 & 68 & 61 & 55 &  -1653 \\
256 & 0.5 & -2.7 & 2.0 & No & 50-50 & 0.1 & 69 & 38 & 17 &  -1659 \\
54 & 0.2 & -2.7 & 1.0 & No & Flat & 0.1 & 70 & 59 & 58 &  -1675 \\
38 & 0.2 & -2.7 & 2.0 & Yes & Flat & 0 & 71 & 49 & 38 &  -1737 \\
214 & 0.2 & -2.7 & 0.25 & No & 50-50 & 0 & 72 & 110 & 121 &  -1771 \\
45 & 0.1 & -2.7 & 0.25 & Yes & Flat & 0 & 73 & 106 & 122 &  -1790 \\
25 & 0.1 & -2.35 & 1.0 & Yes & Flat & 0 & 74 & 94 & 112 &  -1799 \\
86 & 0.2 & -2.7 & 2.0 & Yes & Flat & 0.1 & 75 & 50 & 35 &  -1802 \\
87 & 0.3 & -2.7 & 2.0 & Yes & Flat & 0.1 & 76 & 68 & 53 &  -1815 \\
63 & 0.3 & -2.7 & 2.0 & No & Flat & 0.1 & 77 & 77 & 46 &  -1817 \\
257 & 0.1 & -2.35 & 0.25 & No & 50-50 & 0.1 & 78 & 116 & 135 &  -1843 \\
263 & 0.3 & -2.7 & 0.25 & No & 50-50 & 0.1 & 79 & 105 & 114 &  -1862 \\
6 & 0.2 & -2.7 & 1.0 & No & Flat & 0 & 80 & 65 & 72 &  -1867 \\
251 & 0.3 & -2.35 & 2.0 & No & 50-50 & 0.1 & 81 & 63 & 51 &  -1947 \\
200 & 0.5 & -2.7 & 1.0 & No & 50-50 & 0 & 82 & 83 & 84 &  -1952 \\
15 & 0.3 & -2.7 & 2.0 & No & Flat & 0 & 83 & 70 & 49 &  -1957 \\
275 & 0.3 & -2.35 & 2.0 & Yes & 50-50 & 0.1 & 84 & 67 & 61 &  -1977 \\
242 & 0.2 & -2.35 & 1.0 & No & 50-50 & 0.1 & 85 & 78 & 83 &  -2017 \\
7 & 0.3 & -2.7 & 1.0 & No & Flat & 0 & 86 & 99 & 89 &  -2114 \\
78 & 0.2 & -2.7 & 1.0 & Yes & Flat & 0.1 & 87 & 73 & 67 &  -2125 \\
55 & 0.3 & -2.7 & 1.0 & No & Flat & 0.1 & 88 & 96 & 78 &  -2129 \\
30 & 0.2 & -2.7 & 1.0 & Yes & Flat & 0 & 89 & 82 & 82 &  -2195 \\
224 & 0.5 & -2.7 & 1.0 & Yes & 50-50 & 0 & 90 & 95 & 96 &  -2298 \\
286 & 0.2 & -2.7 & 0.25 & Yes & 50-50 & 0.1 & 91 & 97 & 110 &  -2343 \\
10 & 0.2 & -2.35 & 2.0 & No & Flat & 0 & 92 & 93 & 69 &  -2354 \\
194 & 0.2 & -2.35 & 1.0 & No & 50-50 & 0 & 93 & 102 & 118 &  -2431 \\
272 & 0.5 & -2.7 & 1.0 & Yes & 50-50 & 0.1 & 94 & 88 & 76 &  -2540 \\
266 & 0.2 & -2.35 & 1.0 & Yes & 50-50 & 0.1 & 95 & 90 & 99 &  -2574 \\
82 & 0.2 & -2.35 & 2.0 & Yes & Flat & 0.1 & 96 & 98 & 75 &  -2600 \\
243 & 0.3 & -2.35 & 1.0 & No & 50-50 & 0.1 & 97 & 104 & 97 &  -2604 \\
204 & 0.5 & -2.35 & 2.0 & No & 50-50 & 0 & 98 & 69 & 63 &  -2622 \\
215 & 0.3 & -2.7 & 0.25 & No & 50-50 & 0 & 99 & 132 & 130 &  -2623 \\
35 & 0.3 & -2.35 & 2.0 & Yes & Flat & 0 & 100 & 122 & 100 &  -2748 \\
195 & 0.3 & -2.35 & 1.0 & No & 50-50 & 0 & 101 & 120 & 126 &  -2751 \\
16 & 0.5 & -2.7 & 2.0 & No & Flat & 0 & 102 & 64 & 44 &  -2809 \\
58 & 0.2 & -2.35 & 2.0 & No & Flat & 0.1 & 103 & 92 & 68 &  -2812 \\
31 & 0.3 & -2.7 & 1.0 & Yes & Flat & 0 & 104 & 108 & 101 &  -2822 \\
252 & 0.5 & -2.35 & 2.0 & No & 50-50 & 0.1 & 105 & 62 & 47 &  -2835 \\
34 & 0.2 & -2.35 & 2.0 & Yes & Flat & 0 & 106 & 100 & 80 &  -2840 \\
281 & 0.1 & -2.35 & 0.25 & Yes & 50-50 & 0.1 & 107 & 124 & 134 &  -2849 \\
218 & 0.2 & -2.35 & 1.0 & Yes & 50-50 & 0 & 108 & 107 & 124 &  -2880 \\
11 & 0.3 & -2.35 & 2.0 & No & Flat & 0 & 109 & 113 & 93 &  -2902 \\
276 & 0.5 & -2.35 & 2.0 & Yes & 50-50 & 0.1 & 110 & 85 & 66 &  -2970 \\
40 & 0.5 & -2.7 & 2.0 & Yes & Flat & 0 & 111 & 74 & 54 &  -2993 \\
70 & 0.2 & -2.7 & 0.25 & No & Flat & 0.1 & 112 & 123 & 131 &  -3088 \\
238 & 0.2 & -2.7 & 0.25 & Yes & 50-50 & 0 & 113 & 117 & 128 &  -3097 \\
88 & 0.5 & -2.7 & 2.0 & Yes & Flat & 0.1 & 114 & 75 & 52 &  -3140 \\
22 & 0.2 & -2.7 & 0.25 & No & Flat & 0 & 115 & 137 & 141 &  -3145 \\
79 & 0.3 & -2.7 & 1.0 & Yes & Flat & 0.1 & 116 & 103 & 92 &  -3211 \\
228 & 0.5 & -2.35 & 2.0 & Yes & 50-50 & 0 & 117 & 89 & 77 &  -3354 \\
64 & 0.5 & -2.7 & 2.0 & No & Flat & 0.1 & 118 & 72 & 43 &  -3355 \\
267 & 0.3 & -2.35 & 1.0 & Yes & 50-50 & 0.1 & 119 & 114 & 116 &  -3376 \\
83 & 0.3 & -2.35 & 2.0 & Yes & Flat & 0.1 & 120 & 118 & 95 &  -3642 \\
56 & 0.5 & -2.7 & 1.0 & No & Flat & 0.1 & 121 & 109 & 94 &  -3655 \\
287 & 0.3 & -2.7 & 0.25 & Yes & 50-50 & 0.1 & 122 & 125 & 129 &  -3675 \\
59 & 0.3 & -2.35 & 2.0 & No & Flat & 0.1 & 123 & 112 & 91 &  -3683 \\
258 & 0.2 & -2.35 & 0.25 & No & 50-50 & 0.1 & 124 & 148 & 153 &  -3692 \\
244 & 0.5 & -2.35 & 1.0 & No & 50-50 & 0.1 & 125 & 133 & 120 &  -3721 \\
65 & 0.1 & -2.35 & 0.25 & No & Flat & 0.1 & 126 & 147 & 155 &  -3747 \\
196 & 0.5 & -2.35 & 1.0 & No & 50-50 & 0 & 127 & 142 & 138 &  -3893 \\
239 & 0.3 & -2.7 & 0.25 & Yes & 50-50 & 0 & 128 & 139 & 140 &  -3980 \\
219 & 0.3 & -2.35 & 1.0 & Yes & 50-50 & 0 & 129 & 126 & 133 &  -3993 \\
209 & 0.1 & -2.35 & 0.25 & No & 50-50 & 0 & 130 & 145 & 149 &  -3998 \\
264 & 0.5 & -2.7 & 0.25 & No & 50-50 & 0.1 & 131 & 135 & 137 &  -4004 \\
71 & 0.3 & -2.7 & 0.25 & No & Flat & 0.1 & 132 & 141 & 145 &  -4247 \\
89 & 0.1 & -2.35 & 0.25 & Yes & Flat & 0.1 & 133 & 152 & 156 &  -4269 \\
8 & 0.5 & -2.7 & 1.0 & No & Flat & 0 & 134 & 121 & 106 &  -4304 \\
50 & 0.2 & -2.35 & 1.0 & No & Flat & 0.1 & 135 & 119 & 108 &  -4360 \\
259 & 0.3 & -2.35 & 0.25 & No & 50-50 & 0.1 & 136 & 162 & 166 &  -4454 \\
94 & 0.2 & -2.7 & 0.25 & Yes & Flat & 0.1 & 137 & 136 & 139 &  -4582 \\
216 & 0.5 & -2.7 & 0.25 & No & 50-50 & 0 & 138 & 150 & 144 &  -4608 \\
210 & 0.2 & -2.35 & 0.25 & No & 50-50 & 0 & 139 & 166 & 160 &  -4777 \\
2 & 0.2 & -2.35 & 1.0 & No & Flat & 0 & 140 & 131 & 127 &  -4909 \\
23 & 0.3 & -2.7 & 0.25 & No & Flat & 0 & 141 & 151 & 152 &  -4919 \\
32 & 0.5 & -2.7 & 1.0 & Yes & Flat & 0 & 142 & 134 & 125 &  -4941 \\
233 & 0.1 & -2.35 & 0.25 & Yes & 50-50 & 0 & 143 & 144 & 148 &  -4973 \\
74 & 0.2 & -2.35 & 1.0 & Yes & Flat & 0.1 & 144 & 129 & 123 &  -5056 \\
12 & 0.5 & -2.35 & 2.0 & No & Flat & 0 & 145 & 115 & 90 &  -5068 \\
46 & 0.2 & -2.7 & 0.25 & Yes & Flat & 0 & 146 & 143 & 147 &  -5261 \\
51 & 0.3 & -2.35 & 1.0 & No & Flat & 0.1 & 147 & 140 & 132 &  -5439 \\
26 & 0.2 & -2.35 & 1.0 & Yes & Flat & 0 & 148 & 138 & 136 &  -5439 \\
3 & 0.3 & -2.35 & 1.0 & No & Flat & 0 & 149 & 149 & 142 &  -5475 \\
80 & 0.5 & -2.7 & 1.0 & Yes & Flat & 0.1 & 150 & 130 & 117 &  -5481 \\
60 & 0.5 & -2.35 & 2.0 & No & Flat & 0.1 & 151 & 111 & 86 &  -5491 \\
84 & 0.5 & -2.35 & 2.0 & Yes & Flat & 0.1 & 152 & 127 & 102 &  -5517 \\
41 & 0.1 & -2.35 & 0.25 & Yes & Flat & 0 & 153 & 159 & 164 &  -5557 \\
17 & 0.1 & -2.35 & 0.25 & No & Flat & 0 & 154 & 160 & 165 &  -5648 \\
288 & 0.5 & -2.7 & 0.25 & Yes & 50-50 & 0.1 & 155 & 155 & 158 &  -5707 \\
240 & 0.5 & -2.7 & 0.25 & Yes & 50-50 & 0 & 156 & 161 & 163 &  -5950 \\
268 & 0.5 & -2.35 & 1.0 & Yes & 50-50 & 0.1 & 157 & 146 & 143 &  -6167 \\
220 & 0.5 & -2.35 & 1.0 & Yes & 50-50 & 0 & 158 & 154 & 150 &  -6219 \\
36 & 0.5 & -2.35 & 2.0 & Yes & Flat & 0 & 159 & 128 & 104 &  -6354 \\
211 & 0.3 & -2.35 & 0.25 & No & 50-50 & 0 & 160 & 173 & 170 &  -6397 \\
282 & 0.2 & -2.35 & 0.25 & Yes & 50-50 & 0.1 & 161 & 157 & 157 &  -6626 \\
75 & 0.3 & -2.35 & 1.0 & Yes & Flat & 0.1 & 162 & 153 & 146 &  -6684 \\
18 & 0.2 & -2.35 & 0.25 & No & Flat & 0 & 163 & 175 & 176 &  -7172 \\
47 & 0.3 & -2.7 & 0.25 & Yes & Flat & 0 & 164 & 164 & 162 &  -7192 \\
95 & 0.3 & -2.7 & 0.25 & Yes & Flat & 0.1 & 165 & 158 & 159 &  -7368 \\
66 & 0.2 & -2.35 & 0.25 & No & Flat & 0.1 & 166 & 170 & 174 &  -7542 \\
27 & 0.3 & -2.35 & 1.0 & Yes & Flat & 0 & 167 & 156 & 151 &  -7695 \\
52 & 0.5 & -2.35 & 1.0 & No & Flat & 0.1 & 168 & 165 & 154 &  -7837 \\
4 & 0.5 & -2.35 & 1.0 & No & Flat & 0 & 169 & 167 & 161 &  -7917 \\
234 & 0.2 & -2.35 & 0.25 & Yes & 50-50 & 0 & 170 & 169 & 167 &  -8550 \\
72 & 0.5 & -2.7 & 0.25 & No & Flat & 0.1 & 171 & 163 & 168 &  -8769 \\
24 & 0.5 & -2.7 & 0.25 & No & Flat & 0 & 172 & 168 & 169 &  -9112 \\
283 & 0.3 & -2.35 & 0.25 & Yes & 50-50 & 0.1 & 173 & 171 & 172 &  -9139 \\
67 & 0.3 & -2.35 & 0.25 & No & Flat & 0.1 & 174 & 181 & 183 &  -9148 \\
260 & 0.5 & -2.35 & 0.25 & No & 50-50 & 0.1 & 175 & 177 & 178 &  -9520 \\
212 & 0.5 & -2.35 & 0.25 & No & 50-50 & 0 & 176 & 184 & 181 &  -10327 \\
19 & 0.3 & -2.35 & 0.25 & No & Flat & 0 & 177 & 183 & 184 &  -10550 \\
90 & 0.2 & -2.35 & 0.25 & Yes & Flat & 0.1 & 178 & 176 & 177 &  -11415 \\
48 & 0.5 & -2.7 & 0.25 & Yes & Flat & 0 & 179 & 179 & 182 &  -11490 \\
235 & 0.3 & -2.35 & 0.25 & Yes & 50-50 & 0 & 180 & 180 & 175 &  -11524 \\
96 & 0.5 & -2.7 & 0.25 & Yes & Flat & 0.1 & 181 & 178 & 180 &  -11718 \\
28 & 0.5 & -2.35 & 1.0 & Yes & Flat & 0 & 182 & 174 & 173 &  -12336 \\
76 & 0.5 & -2.35 & 1.0 & Yes & Flat & 0.1 & 183 & 172 & 171 &  -12477 \\
42 & 0.2 & -2.35 & 0.25 & Yes & Flat & 0 & 184 & 182 & 179 &  -12787 \\
284 & 0.5 & -2.35 & 0.25 & Yes & 50-50 & 0.1 & 185 & 185 & 187 &  -15605 \\
236 & 0.5 & -2.35 & 0.25 & Yes & 50-50 & 0 & 186 & 189 & 188 &  -16426 \\
91 & 0.3 & -2.35 & 0.25 & Yes & Flat & 0.1 & 187 & 186 & 185 &  -16623 \\
20 & 0.5 & -2.35 & 0.25 & No & Flat & 0 & 188 & 190 & 190 &  -18594 \\
43 & 0.3 & -2.35 & 0.25 & Yes & Flat & 0 & 189 & 187 & 186 &  -19295 \\
68 & 0.5 & -2.35 & 0.25 & No & Flat & 0.1 & 190 & 188 & 189 &  -19584 \\
44 & 0.5 & -2.35 & 0.25 & Yes & Flat & 0 & 191 & 192 & 192 &  -29890 \\
92 & 0.5 & -2.35 & 0.25 & Yes & Flat & 0.1 & 192 & 191 & 191 &  -30317 \\

\enddata
\tablecomments{Columns are: (1) model number; (2) CE efficiency
$\times$ central concentration; (3) exponent of high-mass power law component
of IMF: Kroupa (2001, $-2.35$) or Kroupa \& Weidner (2003, $-2.7$);
(4) stellar wind strength
parameter; (5) Yes: All possible outcomes of a CE event with
a Hertzsprung gap donor allowed; No: A CE with such a donor star will 
always result to a merger; (6) binary mass ratio distribution: 50-50
means half of the binaries originate in a twin binary distribution
and half in a flat mass ratio distribution; (7) parameter with
which the Hobbs et al. (2005) kick distribution is multiplied for
BHs formed through a SN explosion with negligible ejected mass; 
(8) rank of model based on likelihood value in this paper; 
(9) rank of model based on likelihood value in F13;
(10) rank of model based on likelihood value in T13;
(11) 
natural logarithm of ratio of each global likelihood to
maximum global likelihood in this table:
$\ln(\cal L_{\rm global}/\cal L_{\rm global, max})$.
(The full table is available on-line).}
\end{deluxetable*}

\end{document}